\documentclass[fleqn,usenatbib]{mnras}
\pdfoutput=1
\usepackage{newtxtext,newtxmath}
\usepackage[T1]{fontenc}
\usepackage{threeparttable}
\usepackage{booktabs}
\usepackage{url}
\usepackage{hyperref}
\usepackage{longtable}
\usepackage{graphicx}
\usepackage{listing}
\usepackage[T1]{fontenc}
\usepackage{color}
\DeclareRobustCommand{\VAN}[3]{#2}
\let\VANthebibliography\thebibliography
\def\thebibliography{\DeclareRobustCommand{\VAN}[3]{##3}\VANthebibliography}

\usepackage{graphicx}	
\usepackage{amsmath}	
\usepackage{amssymb}	
\usepackage{lineno}

\title[Gamma-ray bursts with extended emission]{New evidence of multiple channels for the origin of gamma-ray bursts with extended emission}
\author[Q. M. Li et al.]{
	Q. M, Li$^{1}$
	Q. B, Sun$^{2}$\thanks{E-mail: sunqibin@ynao.ac.cn},
	Z. B. Zhang$^{1}$\thanks{E-mail: zbzhang@gzu.edu.cn},
	K. J, Zhang$^{1}$, and
	G, Long$^{2}$
	\\
	$^{1}$Guizhou University, Department of Physics, college of physics, Guizhou university, Guiyang, 550025, People's Republic of China\\
	$^{2}$Yunnan Observatories, Chinese Academy of Sciences, Kunming, 650216, People's Republic of China}
\date{Accepted XXX. Received YYY; in original form ZZZ}

\pubyear{2023}

\begin{document}
	
	\label{firstpage}
	\pagerange{\pageref{firstpage}--\pageref{lastpage}}
	\maketitle
	
	\begin{abstract} 
		Gamma-ray bursts (GRBs) are the most intense explosions in the universe. GRBs with extended emission (GRB EE) constitute a small subclass of GRBs. GRB EE are divided into EE-I GRBs and EE-II GRBs, according to the Amati empirical relationship rather than duration. We test here if these two types of GRB have different origins based on their luminosity function (and formation rate). Therefore, we use Lynden-Bell's $c^-$ method to investigate the LF and FR of GRBs with EE without any assumption. We calculate the formation rate of two types of GRBs. For EE-I GRBs, the fitting function can be written as $\rho (z) \propto {(1 + z)^{ - 0.34{\rm{ \pm 0}}{\rm{.04}}}}$ for $z<2.39$ and $\rho (z) \propto {(1 + z)^{ - 2.34{\rm{ \pm 0}}{\rm{.24}}}}$ for $z>2.39$. The formation rate of EE-II can describe as $\rho (z) \propto {(1 + z)^{ - 1.05{\rm{ \pm 1}}{\rm{.10}}}}$ for $z<0.43$ and $\rho (z) \propto {(1 + z)^{ - 8.44{\rm{ \pm 1}}{\rm{.10}}}}$ for $z>0.43$. The local formation rate are $\rho ({\rm{0) = 0.03\, Gp}}{{\rm{c}}^{{\rm{ - 3}}}}{\rm{y}}{{\rm{r}}^{{\rm{ - 1}}}}$ for some EE-I GRBs and $\rho ({\rm{0) = 0.32 \,Gp}}{{\rm{c}}^{{\rm{ - 3}}}}{\rm{y}}{{\rm{r}}^{{\rm{ - 1}}}}$ for EE-II GRBs. Based on these results, we provide a new evidence that the origins of EE-I GRBs are different from EE-II GRBs from the perspective of event rate. The EE-I GRB could be produced from the death of the massive star, but EE-II GRB bursts may come from other processes that are unrelated to the SFR. Our findings indicate that the GRB with EE could have multiple production channels.
	\end{abstract}
	
	\section{Introduction}
	\label{sect:intro}
	Gamma-ray bursts (GRBs) are the most violent explosions in the universe and emit high-energy radiations which are produced in an ultra-relativistic jet \citep{2007ChJAA...7....1Z,2009ARA&A..47..567G}. In the internal shock model, the inner engine will produce shells with comparable energy but a different Lorentz factors $\Gamma$. The slower shell, which is followed by a faster shell, catches up with it and collides, which can produce the pulse profile observable in most GRBs \citep{1997ApJ...490...92K,1998MNRAS.296..275D,1999PhR...314..575P}.  Traditionally, GRBs are divided into long GRBs (lGRB; $T_{90}$ > 2 s) and short GRBs (sGRB; $T_{90}$ < 2 s) \citep{1993ApJ...413L.101K}. $T_{90}$ is the time interval during which the integrated photon counts accumulate from 5\% to 95\% of the total photon counts in the prompt emission. The measurement of $T_{90}$ is influenced by different instruments.The dividing line of \textit{Swift} GRB is 1 s \citep{2020ApJ...902...40Z,2022ApJ...940....5D} which is very close to the 1.27 s value of GBM GRBs \citep{2014ApJS..211...12G}, and the duration distribution peak at 0.21 and 42.66 for sGRB and lGRB, respectively. lGRBs are generally believed to originate from the death of massive stars. The association between GRB and supernovae provided direct evidence \citep{2003Natur.423..847H,2003ApJ...591L..17S}. Therefore, lGRBs can be seen as a tool for tracking the star formation rate \citep{2016A&A...587A..40P,2015ApJS..218...13Y,2022MNRAS.513.1078D}. sGRB are believed to originate from the merger of binary compact objects. The association of sGRB 170817A with transient gravitational waves supports the idea that some sGRB is produced by binary neutron stars (BNS) system (GW 170817; \citealp{2017PhRvL.119p1101A}). The more comparison of comprehensive properties of GRB associated with supernovae-kilonovae can be found in our recent work \citep{2023MNRAS.tmp.1633L}.

	However, many authors have also reported a special class of GRBs with extended emission (EE), where the EE is defined as a low-intensity burst following the initial main emission \citep{2001A&A...379L..39L,2002ApJ...567.1028C,2005Sci...309.1833B,2006ApJ...643..266N,2020MNRAS.492.3622L}. Currently, there are several popular speculations about the production of EE: (1) spin-down of a strongly magnetized neutron star \citep{2012MNRAS.419.1537B}; (2) a relativistic wind extracting the rotational energy from a protomagnetar \citep{2008MNRAS.385.1455M}; (3) material fallback of the material heated by r-process \citep{2019MNRAS.485.4404D}. 
	Current research shows that the second peak of GRB 000727 occurs seven seconds after the initial peak \citep{2002astro.ph..9219M} .
	Interestingly, EE components can be identified in both lGRBs and sGRBs from the light curve of prompt emission. \citet{2000ApJ...534..248N} suggested an anti-correlation between spectral lag and peak luminosity for lGRB, but this relationship is different for sGRB and lGRB. \citet{2006Natur.444.1044G} found that the lag and luminosity of GRB 060614 with EE belong to the sGRB plane. Recently, the lGRB 211211A was characterized by a main emission (ME) phase (13 s) and an EE phase lasting 55 s \citep{2022Natur.612..223R,2022Natur.612..232Y,2023ApJ...943..146C}. 
	
	The existence of EE makes it confusing to distinguish between lGRB and sGRB by relying only on the criterion of $T_{90}$. There is a general method involving the peak energy in the rest frame ${E_{p,i}}$ and isotropic energy ${E_{iso}}$, named as Amati correlation \citep{2002A&A...390...81A,2005NCimC..28..251A}, which is used to classify different types of GRBs \citep{2014MNRAS.444L..58V,2018PASP..130e4202Z,2020MNRAS.492.1919M,2023MNRAS.tmp.1633L,2023ApJ...950...30Z}. \citet{2013MNRAS.430..163Q} investigated the distribution of the logarithmic deviation of the peak energy in rest frame ($E_{p,i}$). They proposed a statistical classification of GRBs in the $E_{p, i}$ versus $E_{iso}$ plane (Amati GRBs and non-Amati GRBs), in which the Amati type bursts well follow the Amati relation, non-Amati type bursts do not. \citet{2020RAA....20..201Z} divided long/short GRB with EE into two subclasses (EE-I and EE-II) again based on their positions in the ${E_{p, i}} - {E_{iso}}$ plane, and suggested that these two subclasses have different origins by comparing the empirical relationship (e.g., Yonetoku correlation and peak energy distribution, etc..). According to their results, it is more reasonable to classify GRB with EE into types I and II, which motivates us to investigate their progenitors further.

	It is acceptable that lGRBs are associated with the deaths of massive
	stars. Therefore, it is reasonable to use lGRBs to investigate the star formation rate (SFR) \citep{2004ApJ...609..935Y,2009MNRAS.400L..10W,2010ApJ...711..495B,2015ApJS..218...13Y}. The sGRB is thought to be produced from a coalescence of compact objects. \citet{2014ApJ...789...65Y} pointed out that the key to confirming this idea is the formation rate (FR) of sGRB. Theoretically, the sGRB FR will track the SFR with some delay time. The methods to estimate the coalescence rates of binary compact object systems have large difficulties and uncertainties \citep{1995ApJ...454..593L,1999ApJ...526..152F,2002ApJ...572..407B}. If sGRBs are expected to be accompanied by gravitational-wave emission \citep{1992Sci...256..325A,1992ApJ...395L..83N} (e.g., GW 170817 and sGRB 170817A; \citealp{2018GCN.22763....1R}), then the local sGRB FR is directly related to the expected number of GW events in the future. The FR of sGRB has been extensively explored in previous research \citep{2004JCAP...06..007A,2018ApJ...852....1Z,2021ApJ...914L..40D}. Two of the critical properties characterizing the population of GRBs are their formation rate (FR) and luminosity function (LF), which are helpful to profoundly understand the nature of GRBs \citep{2007ApJ...659..958D,2016A&A...587A..40P}. FR and LF respectively represent the number of bursts per unit comoving volume and the relative fraction of bursts with a certain luminosity. The construction of these two distributions, however, requires measuring the redshift. Only some GRBs EE had well redshift measurements a few years ago, such as GRB 060614 \citep{2006Natur.444.1044G}. The luminosity function and formation rate must be urgently studied with the number of GRBs EE with known redshifts increasing.
	
	The previous studies on GRB LF and FR usually used the $\log N - \log P$ distribution \citep{2000astro.ph..4176F,2011MNRAS.416.2174C,2015ApJ...812...33S}. However, the distribution is produced by the luminosity and redshift convolved \citep{2004ApJ...609..935Y}. \citet{2007NewAR..51..539C} pointed out that several selection effects for the observed redshift distribution of GRBs, such as Malmquist bias, observational limit of the satellite, which is the most important selection effect. The \textit{Swift} have a flux limit ${\rm{2}} \times {\rm{1}}{{\rm{0}}^{{\rm{ - 8}}}}erg/c{m^2}/s$. This means that we can't observe a GRB below the flux limit. To correct this selection effect and obtain the intrinsic LF and FR, \cite{1971MNRAS.155...95L} put forward a non-parametric approach named as Lynden-Bell's ${c^ - }$ method to calculate the FR. This method has been applied to many transient phenomena, such as GRB \citep{2004ApJ...609..935Y,2015ApJS..218...13Y,2016ApJ...820...66D,2018ApJ...852....1Z,2021RAA....21..254L,2022MNRAS.513.1078D,2023arXiv230511380D}, Active galactic nucleus \citep{2011ApJ...743..104S,2021ApJ...913..120Z} and FRB \citep{2019JHEAp..23....1D}. The premise of this method is that the luminosity and redshift are independent of each other. Therefore, we should first test for independence between them by using Kendall $\tau$ test method \citep{1992ApJ...399..345E}. Importantly, \citet{2023arXiv230511380D} also used the non-parameters method to investigate the progenitors of low- and high-luminosity GRB samples.

	In this paper, Our main purpose is to study the LF function and FR of the subclass of GRB EE using Lynden-Bell's ${c^ - }$ method to distinguish their origin. Section \ref{sec:data} introduces the sample source and K-correction method. In Section \ref{sec:amati} and Section \ref{sec:mthd}, we describe the Amati relation, Lynden-Bell's ${c^ - }$ and Kendall $\tau$ test method in detail, respectively. In Section \ref{sec:LF and FR}, we present the result of the luminosity function and formation rate of GRB EE (EE-I and EE-II). Finally, Section \ref{sec:Conclusion} presents the conclusion.

	
	\section{Data and K-correction}
	\label{sec:data}
	
\begin{figure*}
		\centering
		\includegraphics[scale=0.32]{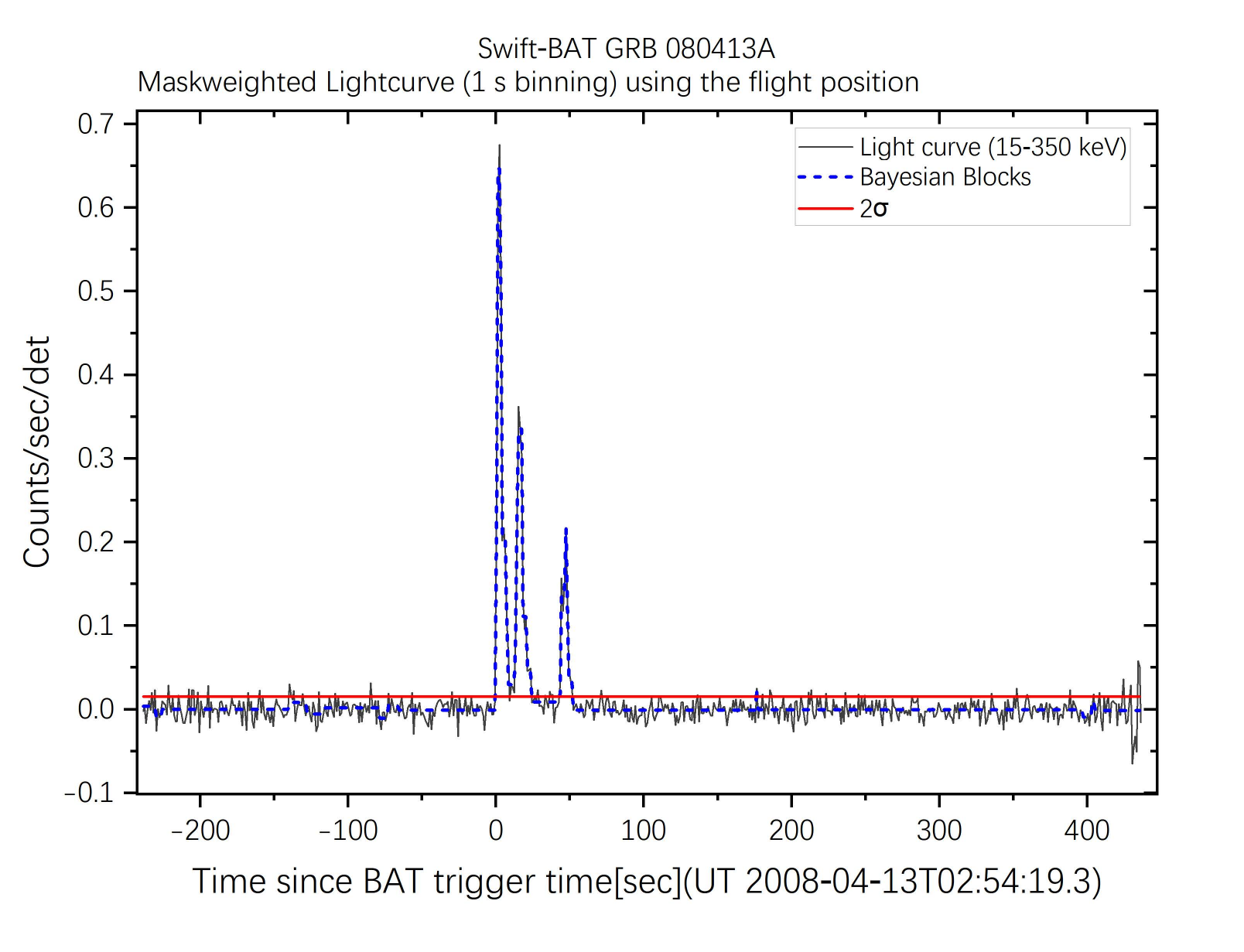}
		\includegraphics[scale=0.32]{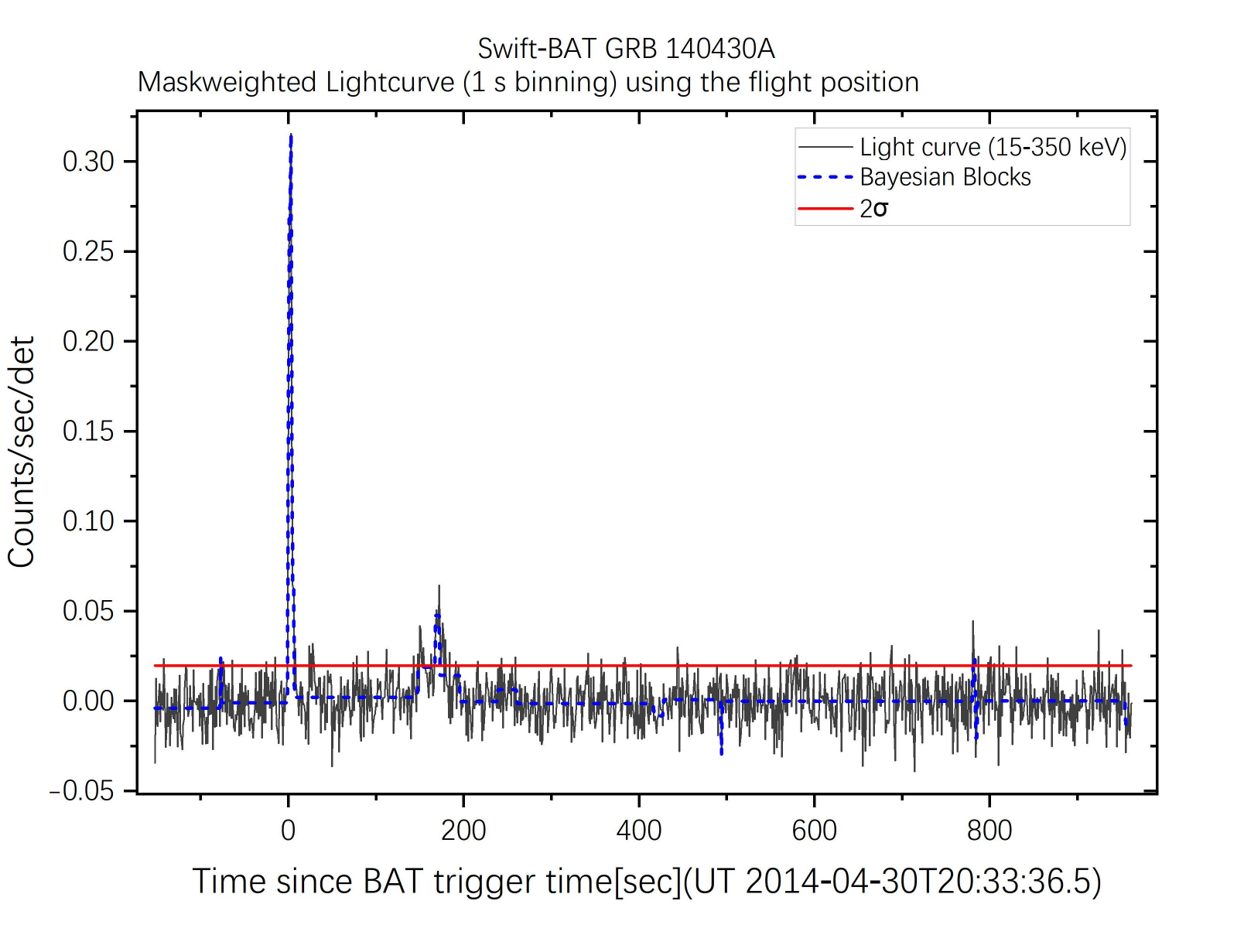}
		\caption{The light curve of prompt emission of GRB 080413A and GRB 140430A in 15-350 keV energy band. The blue dotted line shows the fitted line of the Bayesian block. The red line represents the $2\,\sigma$ line.}    
		\label{fig1}
	\end{figure*}

	\textit{Swift} is a GRB detector with a large wide-field of view that was launched at 17:16 GMT on 20\,th November 2004 \footnote{\url{https://swift.gsfc.nasa.gov/}}. Since 2005, \textit{Swift} has discovered approximately 1600 GRBs, of which 419 GRBs have spectral parameters (estimated redshift, fluence, and peak photon flux). This instrument also provided the light curves in different energy bands, including Channel 1(15-25 keV), Channel 2 (25-50keV), Channel 3 (50-100 keV), Channel 4 (100-350 keV) and Channel 5 (15-350 keV). We combined the ratio of signal to noise ($S/N>2 \sigma$) \citep{2015MNRAS.452..824K,2020RAA....20..201Z,2023MNRAS.tmp.1633L} and Bayesian block method \citep{2010ApJ...717..411N,2013ApJ...764..167S,2014MNRAS.445.2589B,2019ApJ...876...89V} to identify EE components from the light curves of prompt emission. Ultimately, we reserve 80 \textit{Swift} GRBs (60 lGRBs and 20 sGRBs) with well-estimated parameters and EE component. Fig. \ref{fig1} exhibits two typical examples of GRBs with EE component: GRB 080413 and GRB 140430A. In Table \ref{tab:3}, we list the information of GRB EE including GRB name (Column 1), duration $T_{90}$ (Column 2), redshift z (Column 3), low- and high-energy photon index $\alpha$ and $\beta$(Columns 4 and 5), peak energy $E_p$ in observer frame (Column 6), peak flux P (Column 7), fluence $S_{\gamma}$ (Column 8), the energy range $E_{min}-E_{max}$ (Column 9), bolometric peak luminosity $L_p$ and isotropic energy $E_{iso}$ (Columns 10 and 11), and the type of GRB EE (Column 12).

	\begin{figure*}
		\centering
		\includegraphics[scale=0.4]{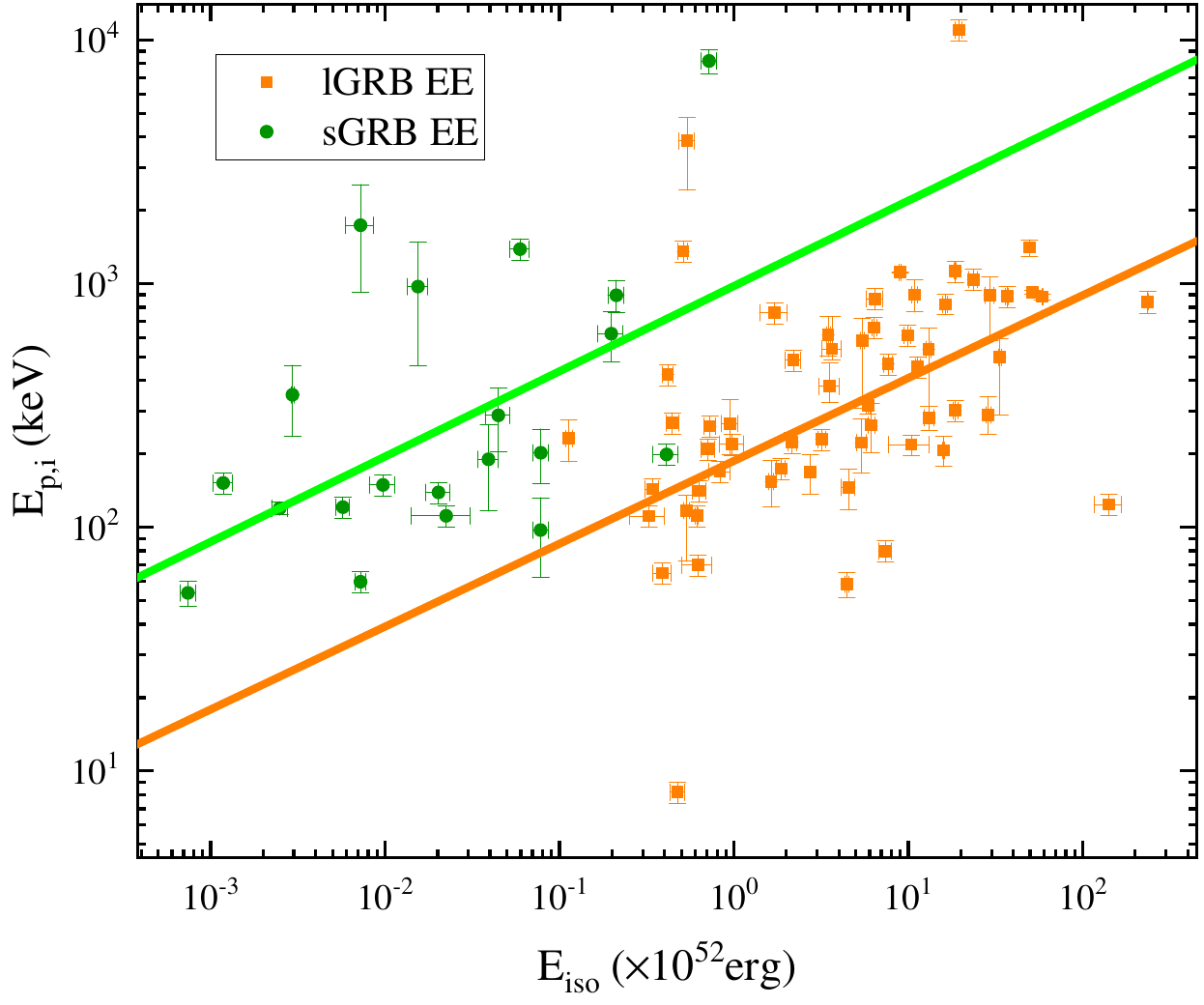}
		\includegraphics[scale=0.4]{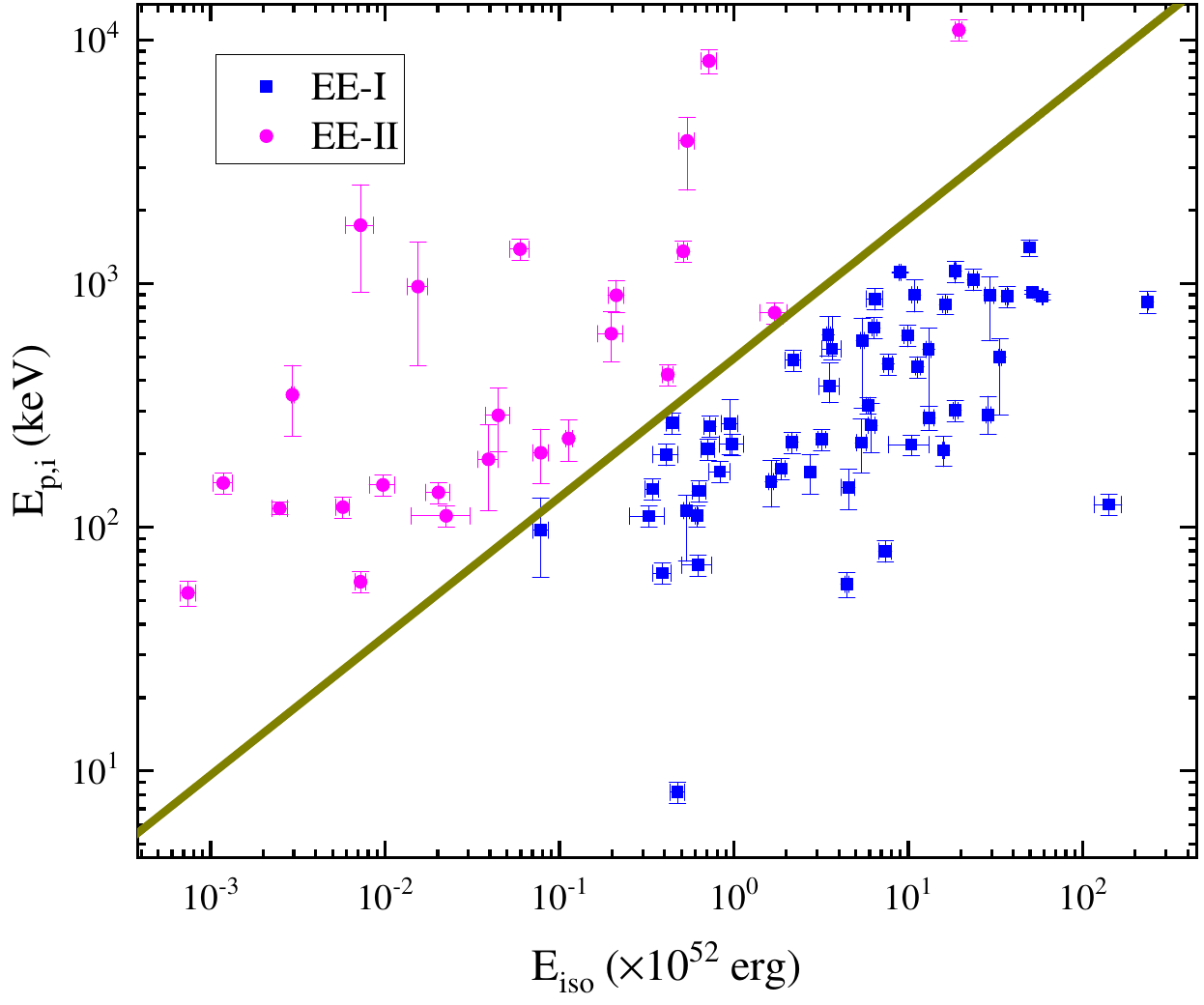}
		\caption{The distribution of 60 lGRB (orange rectangles) and 20 sGRB (green dots) in the $E_{p, i}-E_{iso}$ plane (left panel). The best solid orange and green fitting lines is derived from \citet{2018PASP..130e4202Z} for normal lGRB and sGRB. Right: The distribution of 56 EE-I (blue rectangle) and 24 EE-II GRB (purple dots) in the $E_{p, i}-E_{iso}$ plane (right panel). The brown solid line is taken from \citet{2013MNRAS.430..163Q}.}    
		\label{fig2}
	\end{figure*}
	
	The spectra of GRB are generally fitted by two spectral models, including Band model \citep{1993ApJ...413..281B} and a single/cut-off power law model \citep{2008ApJS..175..179S}. The form of the Band function is as follows:
	\begin{equation}
		\begin{split}
			\Phi (E)=\\ \left\{ {\begin{array}{*{20}{c}}
					{A{{\left( {\frac{E}{{100keV}}} \right)}^\alpha }\exp \left( { - \frac{{(2 + \alpha )E}}{{{E_P}}}} \right),E \le \frac{{(\alpha  - \beta ){E_P}}}{{(2 + \alpha )}}} \hfill\hfill \\
					{A{{\left( {\frac{E}{{100keV}}} \right)}^\beta }\left( {\frac{{(\alpha  - \beta ){E_P}}}{{(2 + \alpha )100keV}}} \right)\exp \left( {\beta {\rm{ - }}\alpha } \right),E \ge \frac{{(\alpha  - \beta ){E_P}}}{{(2 + \alpha )}}}
			\end{array}} \right.
		\end{split}
	\end{equation}
	and the cut off power law can be expressed as
	\begin{equation}
		\Phi (E) = B{\left( {\frac{E}{{50keV}}} \right)^\alpha }\exp \left( { - \frac{{(2 + \alpha )E}}{{{E_P}}}} \right)
	\end{equation}
	Since the peak flux is observed in different energy ranges, we will use the same K-correction method to convert the flux into the $1-10^4$\,keV band to get bolometric luminosities (e.g., \citealp{2015ApJS..218...13Y,2018PASP..130e4202Z}). The bolometric luminosity can be calculated by $L = 4\pi D_L^2(z)P_{bolo}$ or $L = 4\pi D_L^2(z)PK$. $K$ and $P$ are the K-correction factor and the peak flux observed in the energy range, respectively. if P is in units of $erg/c{m^2}/s$, The K can be expressed as
	\begin{equation}
		K = \frac{{\int_{1/(1 + z)}^{10000/(1 + z)} {E\Phi (E)dE} }}{{\int_{{E_{\min }}}^{{E_{\max }}} {E\Phi (E)dE} }}
	\end{equation}
	else if P is in unit of $photon/c{m^2}/s$,  the $P_{bolo}$ can be expressed as
	\begin{equation}
		P_{bolo} =P \frac{{\int_{1/(1 + z)}^{10000/(1 + z)} {E\Phi (E)dE} }}{{\int_{{E_{\min }}}^{{E_{\max }}} {\Phi (E)dE} }}
	\end{equation}
	The value of the flux limit is ${F_{\lim}} = 2.0 \times {10^{ - 8}}erg/c{m^2}/s$ \citep{2015ApJS..218...13Y}. Then, luminosity limit is given as ${L_{\lim}} = 4\pi D_L^2(z){F_{\lim}}$ .
	
	\section{Amati relation}
	\label{sec:amati}
	The duration of a GRB is a key indicator of its physical origin, with lGRBs perhaps associated with the collapse of massive stars and sGRBs with mergers of neutron stars. However, there is a substantial overlap in the properties of both lGRB and sGRB. To date, no other parameter fully distinguishes the origins of these two groups, such as \citet{2023MNRAS.tmp.1633L} who verified that both GRBs associated with supernova/kilonova comply with the Amati relations that match those of long/short GRBs, but two kinds of GRB also have obviously overlapped. Classifying GRBs based on their prompt emission is a useful task, as it has the potential to quickly identify the possible properties of their progenitor and plan the most comprehensive follow-up actions within a few minutes of detecting GRB. However, the situation in this field is puzzling due to the complexity of the GRB light curves and the diversity of possible progenitors (the combination of different types of compact stars, collapsar with or without short-lived active neutron stars, etc.). Fortunately, the sample size of GRB is large enough to establish correlation relationships for classification. The ${E_{p, i}-E_{iso}}$ relation proposed by \citet{2002A&A...390...81A} is a universal method to classify the GRB into lGRB and sGRB. \citet{2006MNRAS.372..233A} implied the sGRB is an outlier of this correlation, so sGRB could have its own Amati relation. They found that the slope of lGRBs is 0.5 (see also \citealp{2009ApJ...703.1696Z}), showing that it can be a powerful tool for discriminating different classes of GRBs and understanding their nature and differences.  \citet{2018PASP..130e4202Z} established the $E_{p, i}-E_{iso}$ relationship for lGRB and sGRB, where the lGRB and sGRB distributions are in different locations, although the slopes are consistent. This claim has been challenged by some authors \citep{2005ApJ...627..319B,2009ApJ...704.1405K,2013A&A...557A.100H}, suggesting that the relationship is the result of selection and instrumental effects, but some authors have argued that these effects are relatively small \citep{2013IJMPD..2230028A,2017A&A...598A.112D,2017A&A...598A.113D}.\citet{2020RAA....20..201Z} found that reclassifying the GRB with EE into EE-I and EE-II types can result in a tighter correlation. This classification is similar to that of \citet{2013MNRAS.430..163Q}, who proposed that the GRB in the Amati plane could be divided into two groups: Amati GRB and no-Amati GRB based on the logarithmic deviation of the $E_p$. Therefore, according to their empirical relationship, we divide our samples into two categories (see Fig. \ref{fig2} right panel): Amati GRB (renamed as type EE-I) and no-Amati GRB (renamed as type EE-II), defined as GRBs that are located below and above the empirical relationship ($E_{p,r, pre }= 493 E_{iso }^{0.57}$), and the $E_{iso}$ is in units of $10^{52} erg$.
	
	The isotropic bolometric energy can be expressed as  $E = 4\pi D_L^2(z)S_{\gamma}K/(1+z)$, and the peak energy in the rest frame can be calculated by $E_{p, i}=E_p(1+z)$. Fig. \ref{fig2}a shows the $E_{p, i}-E_{iso}$  locations of lGRB and sGRB with EE. The best fits the Amati relations of short and long bursts are taken from \citet{2018PASP..130e4202Z}. We redivide these EE GRB into type EE-I and EE-II as Fig. \ref{fig2}b, according to the empirical relationship proposed by \citet{2013MNRAS.430..163Q}. Out of 56 EE-I types, 54 belong to lGRB (94\%), and Out of 24 EE-II types, 18 belong to sGRB (75\%), which indicates that most EE-I type bursts are lGRB and most non EE-II type bursts are of sGRBs.

	\section{LYNDEN-BELL'S ${c^ - }$ METHOD AND NON-PARAMETERIC TEST METHOD}
	\label{sec:mthd}
	The Lynden-Bell ${c^ - }$ method is an effective non-parametric method to analyze the distribution of the bolometric luminosity/energy and redshift of the astronomical objects with the truncated data sample (eg., \citealp{2004ApJ...609..935Y,2015ApJS..218...13Y,2016ApJ...820...66D,2021RAA....21..254L,2022MNRAS.513.1078D}), Active galactic nucleus \citep{2011ApJ...743..104S,2021ApJ...913..120Z} and FRB \citep{2019JHEAp..23....1D}. This work also uses this method to study LF and FR.
	
	If the parameters $L$ and $z$ are independent, the distribution of LF and redshift $\Psi (L,z) = {\psi _z}(L)\varphi (z)$ can be written as $\Psi (L,z) = \psi _z (L)\varphi (z)$. The ${\psi _z}(L)$ is LF at redshift $z$. A function $g(z) = {(1 + z)^k}$ can remove the effect of luminosity evolution. Then the $L$ will transform into ${L_0} = L/g{(z)}$. Therefore, $\Psi (L,z)$ can write as $\psi ({L_0})\varphi (z)$.
	
	\begin{figure}
		\centering
		\includegraphics[scale=0.4]{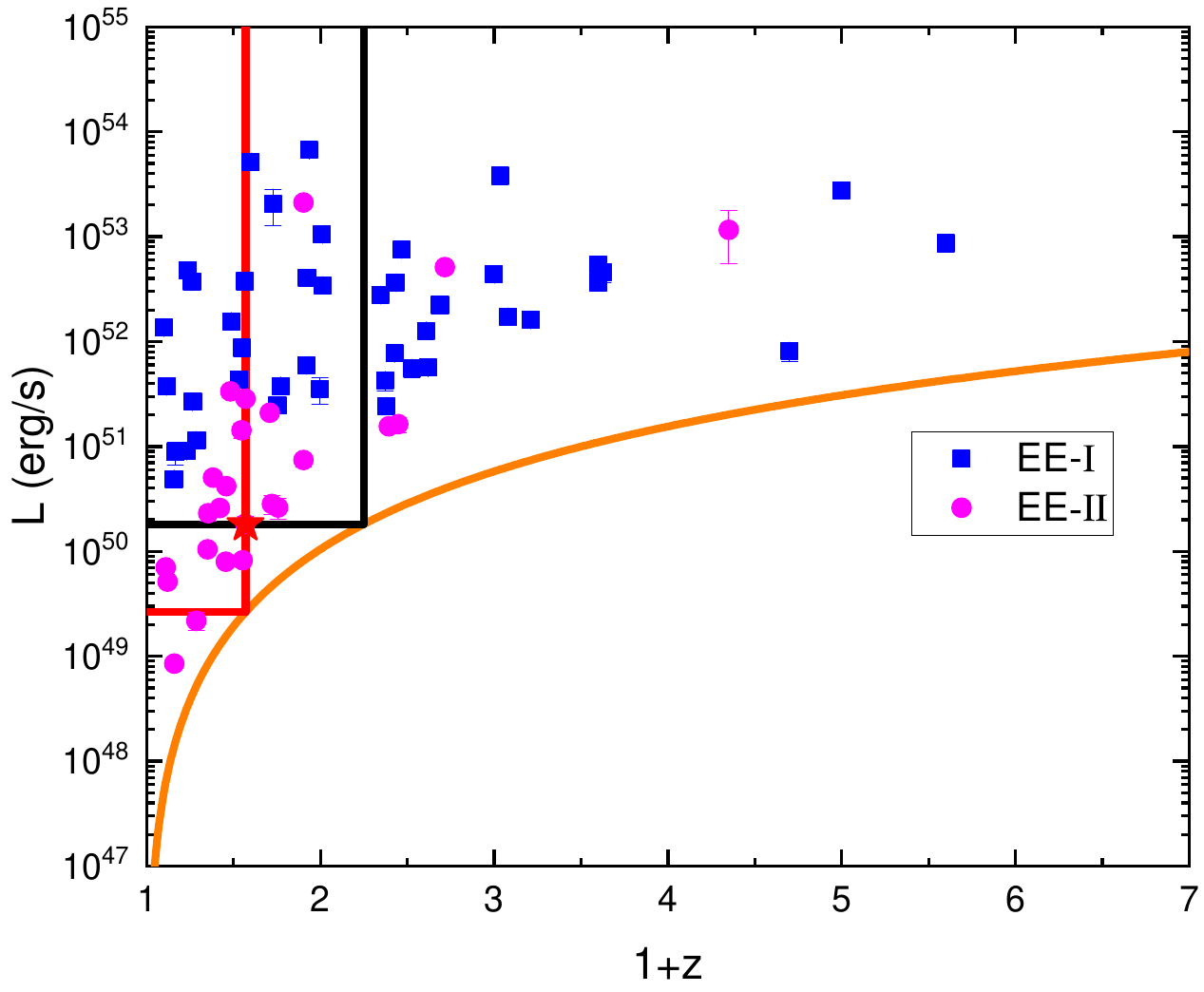}
		\caption{The distribution of luminosity and redshift in the $L-z$ plane. The blue squares and purple dots represent the EE-I and EE-II GRBs, respectively. The flux limit is $2 \times {10^{ - 8}}erg/c{m^2}/s$.}
		\label{fig3}
	\end{figure}

	We use the non-parametric test method raised by \citet{1992ApJ...399..345E} to derive the evolution function $g(z)$. In the $(L, z)$ plane as shown in Fig. \ref{fig3}, for the ith point $({L_i},{z_i})$, we can define $J_i$ as
	\begin{equation}
		{J_i} = \{ j|{L_j} \ge {L_i},z_j \le z_i^{\max }\}
	\end{equation}
	where $L_i$ is the luminosity of the $i$th GRB EE and $z_i^{\max }$ is the maximum redshift at which a GRB EE (EE-I and EE-II) with the luminosity $L_i$ can be detected by $\textit{Swift}$ detector. This range is shown as a black rectangle in Fig. \ref{fig3}. The number included in this range is $n_i$, and the $N_i$ is defined as $n_i-1$. which means take $i$th out. and the $J_{_i}^1$ also can be defined as
	\begin{equation}
		J_{_i}^1 = \{ j|{L_j} \ge L_i^{\min },{z_j} \le {z_i}\}
	\end{equation}
	where $L_i^{\min }$ is the limit luminosity at the redshift $z_i$. This range is shown as the red rectangle in Fig. \ref{fig3}. The number included in this region is $M_i$.

	\begin{figure}
		\centering
		\includegraphics[scale=0.4]{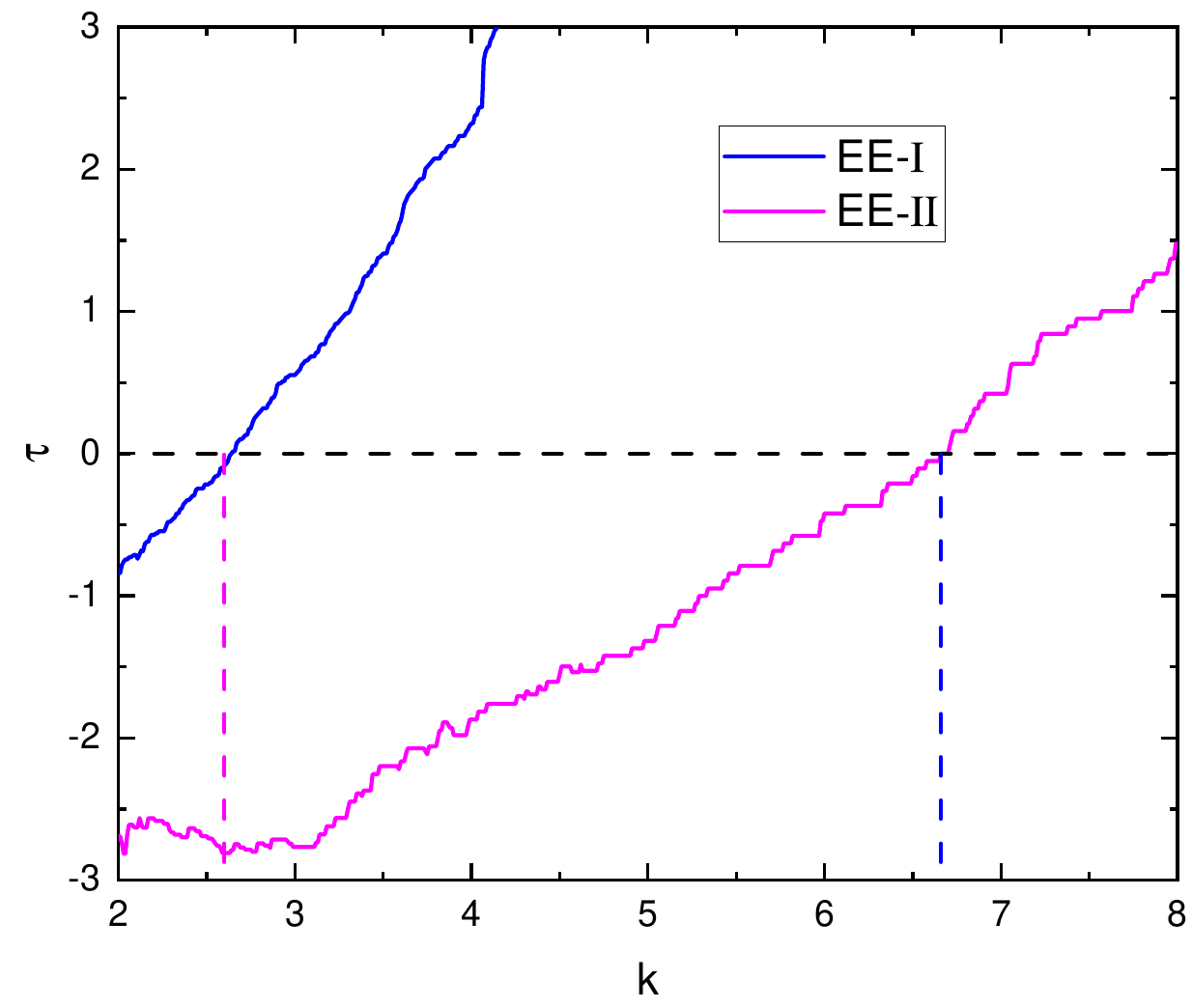}
		\caption{The evolution of $\tau$ as $k$. The blue and purple dotted lines represent the $k$ value when the $\tau =0$ for EE-I and EE-II GRBs, respectively.}
		\label{fig4}
	\end{figure}
	
	In the black rectangle, $R_i$ is defined as the events number that have redshift $z$ less than $z_i$. $R_i$ should be uniformly distributed between 1 and $n_i$ based on the fact that $L$ and $z$ are independent. The Kendall $\tau$ test statistic is \citep{1992ApJ...399..345E}
	\begin{equation}
		\tau  = \sum\limits_i {\frac{{({R_i} - {E_i})}}{{\sqrt {{V_i}} }}}
	\end{equation}
	where ${{E_i} = \frac{{1 + {n_i}}}{2}}$ and ${{V_i} = \frac{{n_i^2 - 1}}{{12}}}$ are respectively the expected mean and variance of $R_i$. $\tau$ will be zero if the size of the sample of ${R_i} \le {E_i}$ is equal to the size of the sample with ${R_i} \ge {E_i}$. After we find the function form of $g(z)$, the effection of luminosity evolution can be removed by transforming $L$ into $L_0$.
	
	$L$ and $z$ are independent on each other until the test statistic $\tau$ is zero by changing the value of k. We show how $\tau$ changes with varying $k$. The $k$ value is 2.64 and 6.66 for EE-I and EE-II in Fig. \ref{fig4}.

	\begin{figure*}
		\centering
		\includegraphics[scale=0.4]{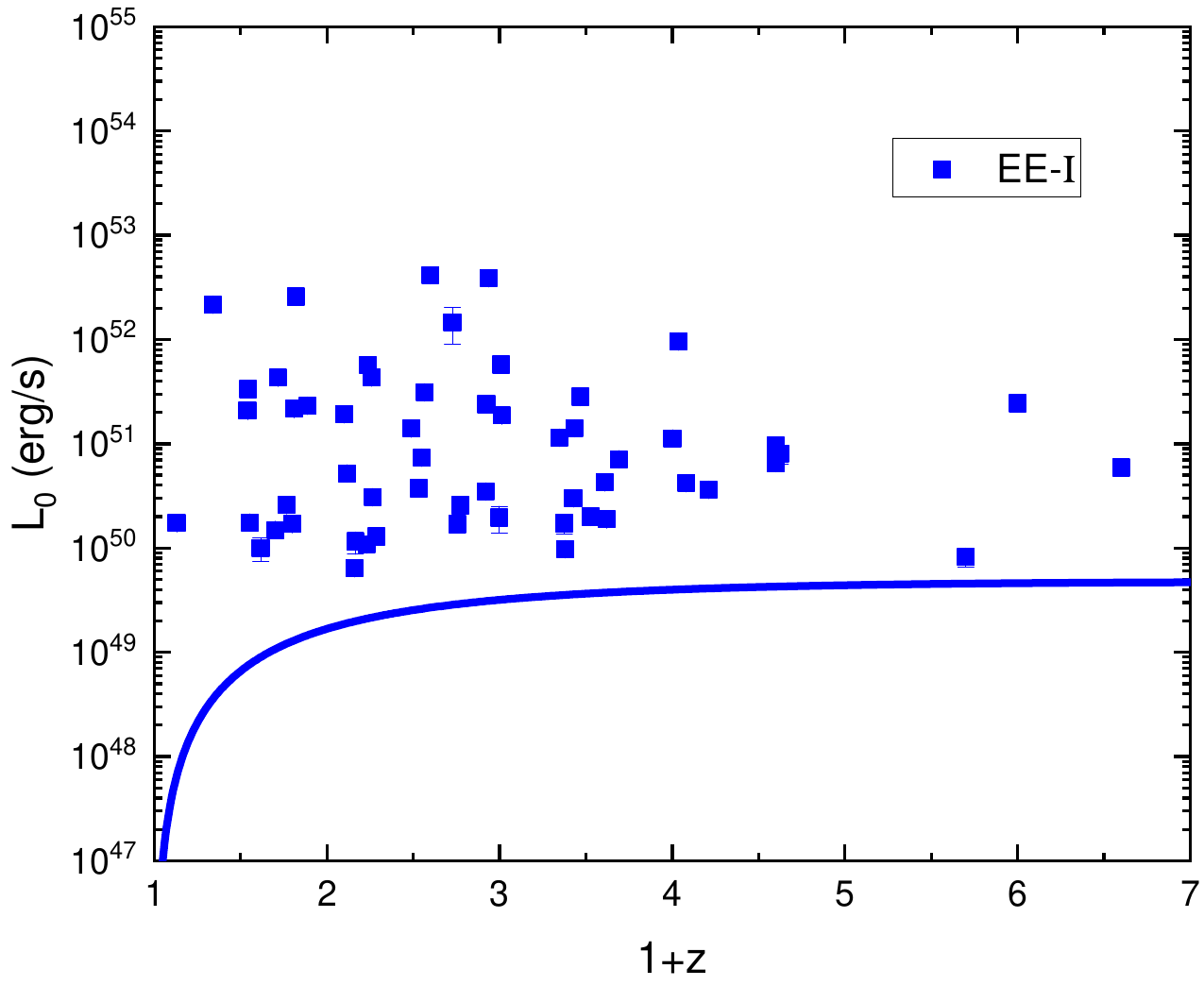}
		\includegraphics[scale=0.4]{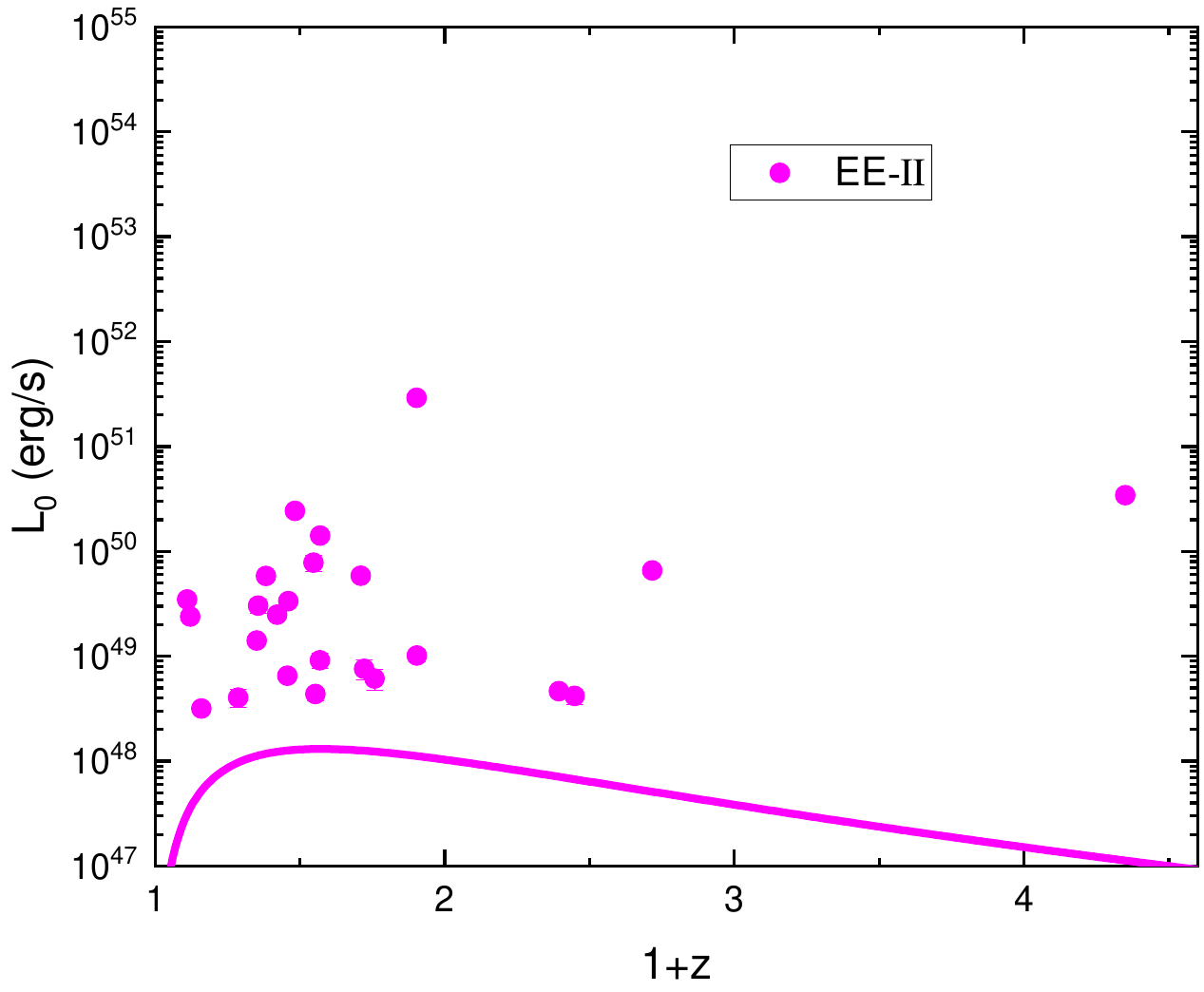}
		\caption{Non-evolving luminosity ${{\rm{L}}_{\rm{0}}}{\rm{ = L/(1 + z}}{{\rm{)}}^{k}}$ of 60 GRBs EE (56 EE-I GRBs and 24 EE-II GRBs) above the truncation line. The k value is 2.64 and 6.66 for EE-I GRB and EE-II GRB, respectively.}
		\label{fig5}
	\end{figure*}
	
	Therefore, the non-evolving luminosity can be written as ${L_0} = L/{(1 + z)^{k}}$ in Fig. \ref{fig5}. We can use a non-parametric method to derive the local cumulative LF distribution from the following equation \citep{1971MNRAS.155...95L,1992ApJ...399..345E}
	\begin{equation}
		\Psi ({L_{0i}}) = \prod\limits_{j < i} {\left( {{\rm{1 + }}\frac{1}{{{N_j}}}} \right)}
	\end{equation}
	and the cumulative number distribution can be obtained from
	\begin{equation}
		\phi ({z_i}) = \prod\limits_{j < i} {\left( {{\rm{1 + }}\frac{1}{{{M_j}}}} \right)}
	\end{equation}
	
	\begin{figure}                           
		\centering
		\includegraphics[scale=0.4]{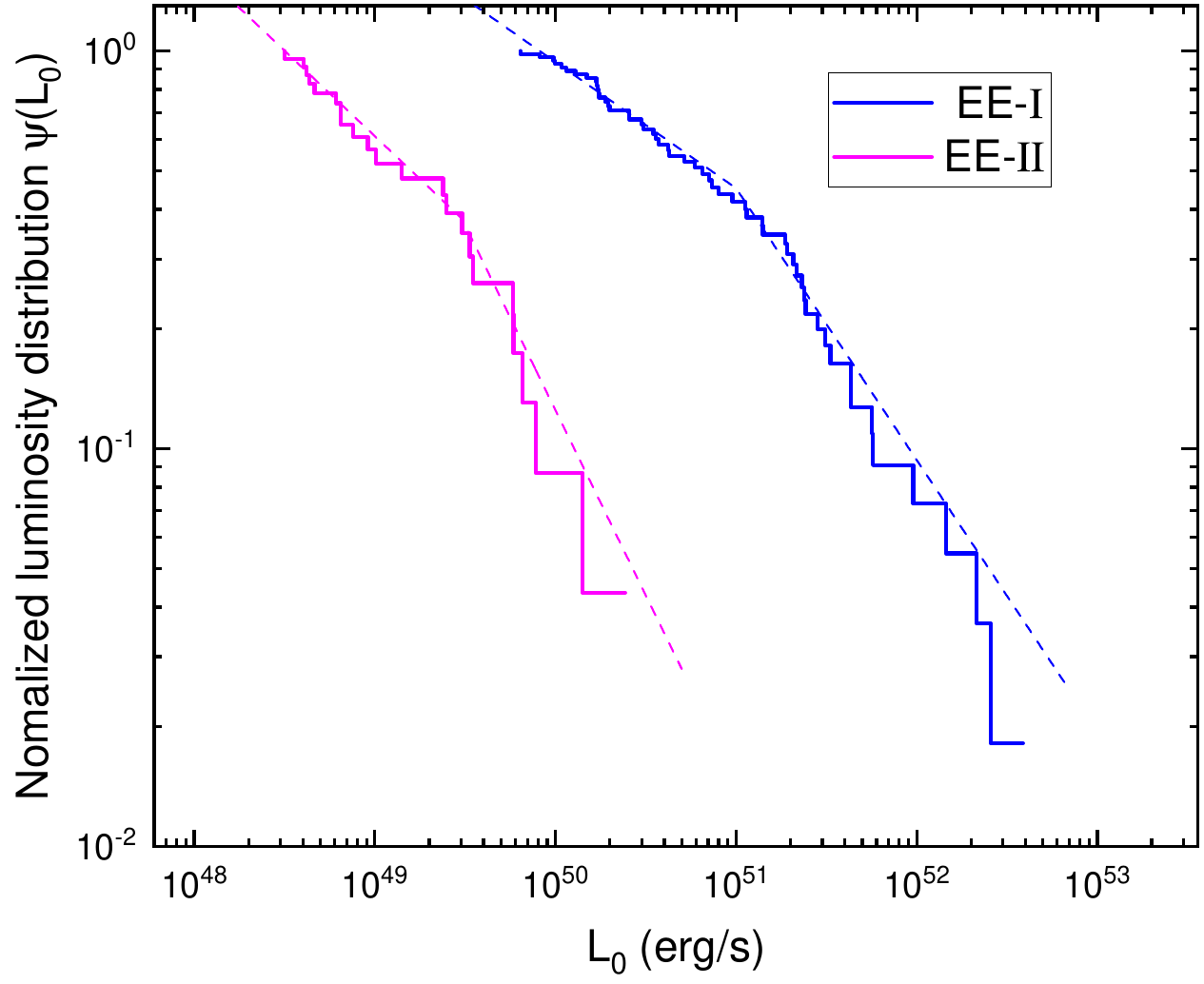}
		\caption{The distribution of local luminosity function for 56 EE-I GRBs and 24 EE-II GRBs). This fitted form can be expressed as $\psi ({L_0}) \propto L_0^{ - 0.34{\rm{ \pm 0}}{\rm{.01}}}$ for the dim segment and $\psi ({L_0}) \propto L_0^{ - 0.67{\rm{ \pm 0}}{\rm{.02}}}$ for the bright segment with EE-I GRB, the broken point are ${L_0^b = 1.00 \times {{10}^{51}}erg/s}$. For EE-II GRBs, the form can be expressed as $\psi ({L_0}) \propto L_0^{ -0.42{\rm{ \pm 0}}{\rm{.02}}}$ for the dim segment and $\psi ({L_0}) \propto L_0^{ - 0.93{\rm{ \pm 0}}{\rm{.01}}}$  for the bright segment with EE-I GRB the broken point is ${L_0^b = 3.03 \times {{10}^{49}}erg/s}$.}
		\label{fig6}
	\end{figure}
	
	Next, the FR can be calculated by
	\begin{equation}
		\rho (z) = \frac{{d\phi (z)}}{{dz}}(1 + z){\left( {\frac{{dV(z)}}{{dz}}} \right)^{ - 1}}
		\label{equ:11}
	\end{equation}
	the ${\frac{{dV(z)}}{{dz}}}$ is the differential comoving volume, which can be expressed as
	\begin{equation}
		\begin{split}
			\frac{{dV(z)}}{{dz}} = 4\pi {\left( {\frac{c}{{{H_0}}}} \right)^3}{\left( {\int_0^z {\frac{{dz}}{{\sqrt {1 - {\Omega _m} + {\Omega _m}{{(1 + z)}^3}} }}} } \right)^{\rm{2}}}\\
			\left( {\frac{1}{{\sqrt {1 - {\Omega _m} + {\Omega _m}{{(1 + z)}^3}} }}} \right) \hfill\hfill
		\end{split}
		\label{equ:12}
	\end{equation}
	
	and the expected number of GRBs can be estimated by \citep{2019MNRAS.488.4607L}
	\begin{equation} 
		\label{equ:N}
		N_{\rm exp} = \frac{\Delta \Omega T}{4\pi} \int^{z_{\rm max}}_{0}  \frac{\rho(z)}{1+z}\frac{dV(z)}{dz} dz \int^{\textit{L}_{\rm max}}_{\rm max[\textit{L}_{\rm min}, \textit{L}_{\rm lim}(z)]} \psi(L)dL
	\end{equation}
	The \textit{Swift} instrument has been running for approximately T=19 years. the field of view of this telescope is $\Omega=1.33sr$ \citep{2015ApJ...812...33S}.
	
	\begin{figure}
		\centering
		\includegraphics[scale=0.4]{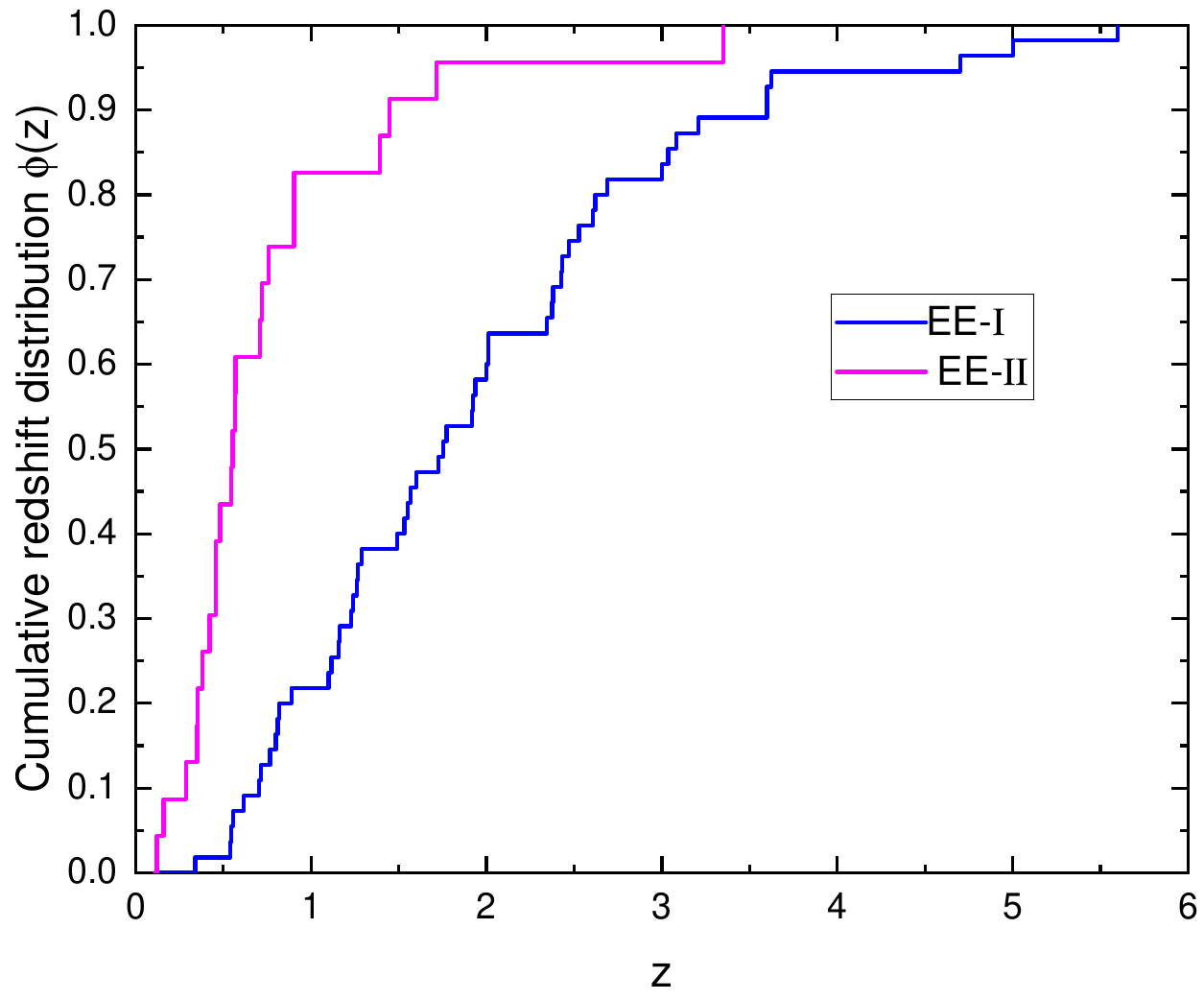}
		\caption{The normalized cumulative distribution of redshift for 56 EE-I and 24 EE-II GRBs).}    
		\label{fig7}
	\end{figure}
	
	\section{Result}
	\label{sec:LF and FR}
	In this part, We present the LF and FR of subclasses of GRBs EE, respectively.
	
	\subsection{Luminosity function}
	Fig. \ref{fig6} shows the distribution of normalized LF $\psi ({L_0})$. Using the broken power law, we can fit this curves to obtain the forms of LF for the dim segment and bright segment for EE-I GRB as
	
	\begin{equation}
		\psi ({L_0}) \propto \left\{ {\begin{array}{*{20}{c}}
				{L_0^{ - 0.34{\rm{ \pm 0}}{\rm{.01}}},{L_0} < L_0^b}\\
				{L_0^{ - 0.{\rm{67 \pm 0}}{\rm{.02}}},{L_0} > L_0^b}
		\end{array}} \right.
	\end{equation}
	and for EE-II as
	\begin{equation}
		\psi ({L_0}) \propto \left\{ {\begin{array}{*{20}{c}}
				{L_0^{ - 0.43{\rm{ \pm 0}}{\rm{.02}}},{L_0} < L_0^b}\\
				{L_0^{ - 0.{\rm{93 \pm 0}}{\rm{.01}}},{L_0} > L_0^b}
		\end{array}} \right.
	\end{equation}
	
	where break point ${L_0^b = 3.03 \times {{10}^{49}}erg/s}$ for EE-II GRBs, which is small two order than the break luminosity ${L_0^b = 1.00 \times {{10}^{51}}erg/s}$ of EE-I GRBs. \citet{2015ApJS..218...13Y} described the cumulatively luminosity distribution by a broken power law function with $\alpha=-0.14\pm0.02$, $\beta=-0.7\pm0.03$, ${L_0^b = 1.43 \times {{10}^{51}}erg/s}$ for 127 lGRBs. \citet{2016A&A...587A..40P} estimated the luminosity distribution of complete 99 lGRBs with $\alpha=-1.32\pm0.21$, $\beta=-1.84\pm0.24$ and ${L_0^b = 2.82 \times {{10}^{51}}erg/s}$. \citet{2015MNRAS.448.3026W} used sGRB, which originated from non-corecollapsars to estimate the LF and FR, and they acquire the breaking point as ${L^b = 2.0 \times {{10}^{52}}erg/s}$ with power law indices of 0.95 and 2.0 for the dim and bright segments, respectively. \citet{2021RAA....21..254L} used 324 \textit{Fermi} sGRB to derive the break point as $\alpha=-0.45\pm0.01$, $\beta=-1.11\pm0.01$, with the break luminosity ${L_0^b = 4.92 \times {{10}^{49}}erg/s}$. It is worth noting that this result is roughly consistent with our work. The break luminosity of EE-I GRBs is larger by two orders than EE-II GRBs, similar with lGRB to sGRB after removing the luminosity evolution with redshift. We must emphasize that the luminosity function only presents the local distribution at $z=0$. Therefore, the LF at redshift z will be rewritten as ${\psi _z}(L) = \psi (L/g(z)) = \psi (L/{(1 + z)^{k}})$. The break point is $L_z^b = L_0^b{(1 + z)^{k}}$ at z.
	\subsection{Formation rate}
	\begin{figure}                               
		\centering
		\includegraphics[scale=0.4]{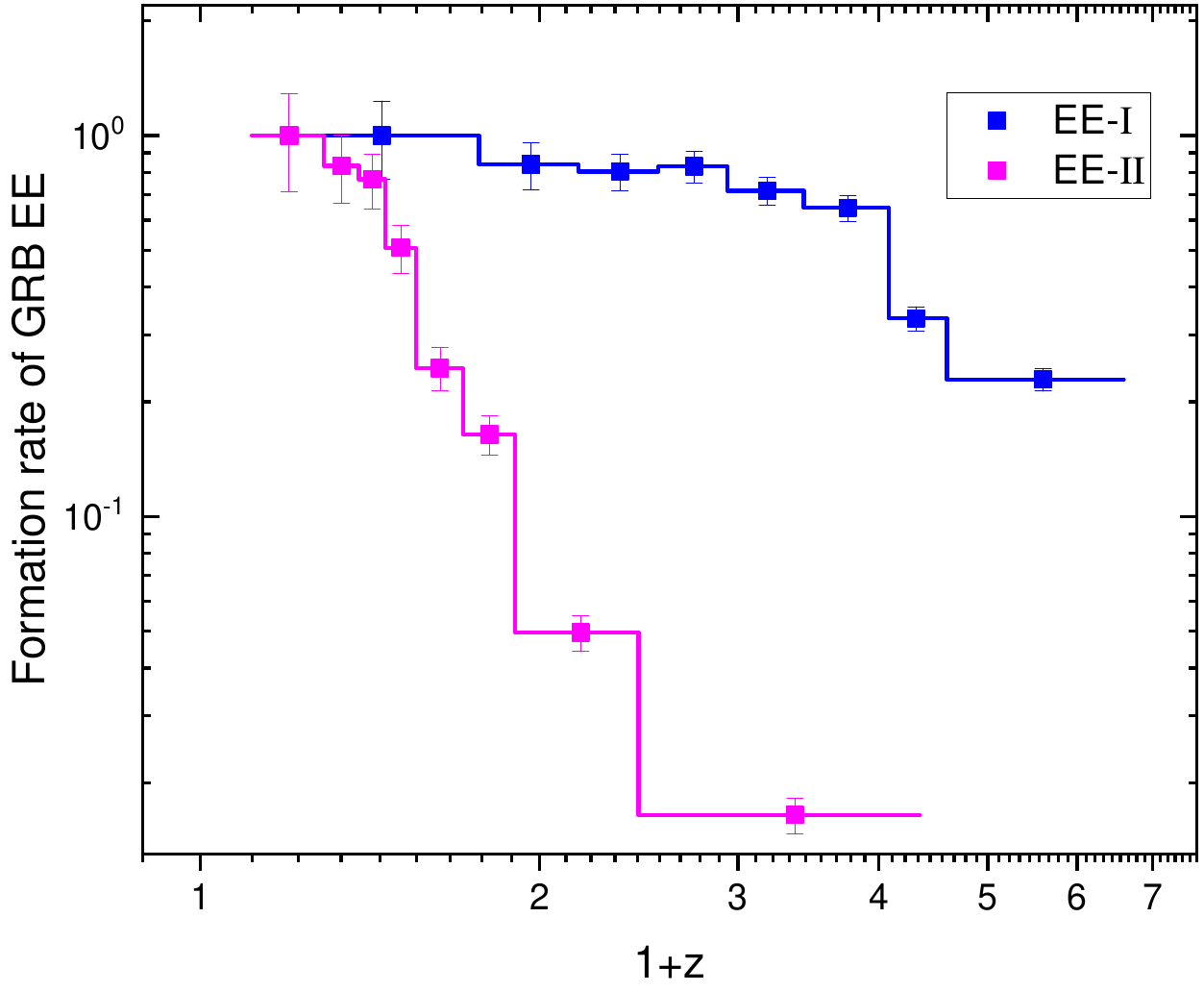}
		\caption{The formation rate for 56 EE-I and 24 EE-II GRBs. The fit function can be write and $\rho (z) \propto {(1 + z)^{ - 0.34{\rm{ \pm 0}}{\rm{.04}}}}$ for $z<2.39$ and $\rho (z) \propto {(1 + z)^{ - 2.34{\rm{ \pm 0}}{\rm{.24}}}}$ for $z>2.39$ for EE-I GRBs. The formation rate of EE-II can describe as $\rho (z) \propto {(1 + z)^{ -1.05{\rm{ \pm 0}}{\rm{.03}}}}$ for $z<0.43$ and $\rho (z) \propto {(1 + z)^{ -8.44{\rm{ \pm 1}}{\rm{.10}}}}$ for $z>0.43$.}
		\label{fig8}
	\end{figure}

	Fig. \ref{fig7} shows the normalized cumulative distribution of redshift $\phi (z)$. From Eq.\ref{equ:11}, we can calculate the FR of GRB EE. Firstly, the differential cumulative redshift distribution form should be derived. In Fig. \ref{fig8}, the blue and purple stepwise line is the FR of GRBs EE. It is obvious that EE-II GRBs has kept a decreasing trend. The error bars is calculate by the GRBs number of black or red rectangle, $M_j$ and $N_j$. The final error of the FR obtained through the error transfer formula.The error bar gives a 1 $\sigma$ poisson error \citep{1986ApJ...303..336G}. We also fit the different segments using the broken power law function. The forms of FR for EE-I GRB can be expressed as

	\begin{equation}\label{eq:z}
		\rho(z)=\rho_{\rm GRB}(0) \left\{
		\begin{aligned}
			{{{(1 + z)}^{ - 0.34{\rm{ \pm 0}}{\rm{.04}}}},z < {z_0}}\\
			{{{(1 + z)}^{ - 2.34{\rm{ \pm 0}}{\rm{.24}}}},z > {z_0}}
		\end{aligned}
		\right.
	\end{equation}
	and for EE-II GRB as
	\begin{equation}\label{eq:z}
		\rho(z)=\rho_{\rm GRB}(0) \left\{
		\begin{aligned}
			{{{(1 + z)}^{ - 1.05{\rm{ \pm 0}}{\rm{.03}}}},z < {z_0}}\\
			{{{(1 + z)}^{ - 8.44{\rm{ \pm 1}}{\rm{.10}}}},z > {z_0}}
		\end{aligned}
		\right.
	\end{equation}

	where break point are ${z_0}=2.39$ for EE-I GRB and ${z_0}=0.43$ for EE-II GRB, the local FR are  $\rho ({\rm{0) =  0.03\, Gp}}{{\rm{c}}^{{\rm{ - 3}}}}{\rm{y}}{{\rm{r}}^{{\rm{ - 1}}}}$ for EE-I GRBs and $\rho ({\rm{0) = 0.32 \, Gp}}{{\rm{c}}^{{\rm{ - 3}}}}{\rm{y}}{{\rm{r}}^{{\rm{ - 1}}}}$ for EE-II GRBs according equation \ref{equ:N}. \citet{2015ApJS..218...13Y} found that the power law index of FR of 127 \textit{Swift} lGRB are $0.04\pm0.94$ for $z<1$, $-0.94\pm0.11$ for $1<z<4$ and  $-4.36\pm0.48$ for $z>4$, and the local formation rate is  $\rho ({\rm{0) = 7.3\pm 2.7\, Gp}}{{\rm{c}}^{{\rm{ - 3}}}}{\rm{y}}{{\rm{r}}^{{\rm{ - 1}}}}$. Importantly, they stress that the GRB rate exceeds the SFR at $z<1$. However, \citet{2016A&A...587A..40P} found that the GRB rate increases to $z=2$ using the complete lGRB. They suggested that the low redshift excess was caused by an incomplete sample, and their results showed that the rate of GRB had a similar trend with SFR. \citet{2021A&A...649A.166P} used models to fit the three observational parameters constrained, including intensity, spectrum, and redshift, and they noted the distribution of LGRB rate follows the shape of cosmic SFR with $a=1.35\pm0.10$, $b=-0.18\pm0.02$, the break redshift $z_m=2.2 \pm 0.10$ if an LF does not evolve with redshift ($k_{evol} = 0$), and $\rho ({\rm{0) = 0.77\pm0.05\, Gp}}{{\rm{c}}^{{\rm{ - 3}}}}{\rm{y}}{{\rm{r}}^{{\rm{ - 1}}}}$. The local formation rate of  $\rho ({\rm{0) =  0.03\, Gp}}{{\rm{c}}^{{\rm{ - 3}}}}{\rm{y}}{{\rm{r}}^{{\rm{ - 1}}}}$ for EE-I GRBs appears very low, compared to the rate of lGRBs, which is of the order of 1 Gpc$^{-3}$ yr$ ^{-1} $. This could be due to the small fraction of Type-I GRBs with extended emission.
	
	Based on empirical relationships proposed by \citet{2004ApJ...609..935Y}, \citet{2021RAA....21..254L} obtained the formation rate of \textit{Fermi} sGRB that can be described as $a=-4.02\pm1.34$, $b=4.93\pm0.30$, the break redshift $z_m=0.4 \pm 0.10$. They also estimate the local sGRB FR is  $\rho ({\rm{0) = 17.43\pm0.12\, Gp}}{{\rm{c}}^{{\rm{ - 3}}}}{\rm{y}}{{\rm{r}}^{{\rm{ - 1}}}}$.

	\begin{figure}                               
		\centering
		\includegraphics[scale=0.4]{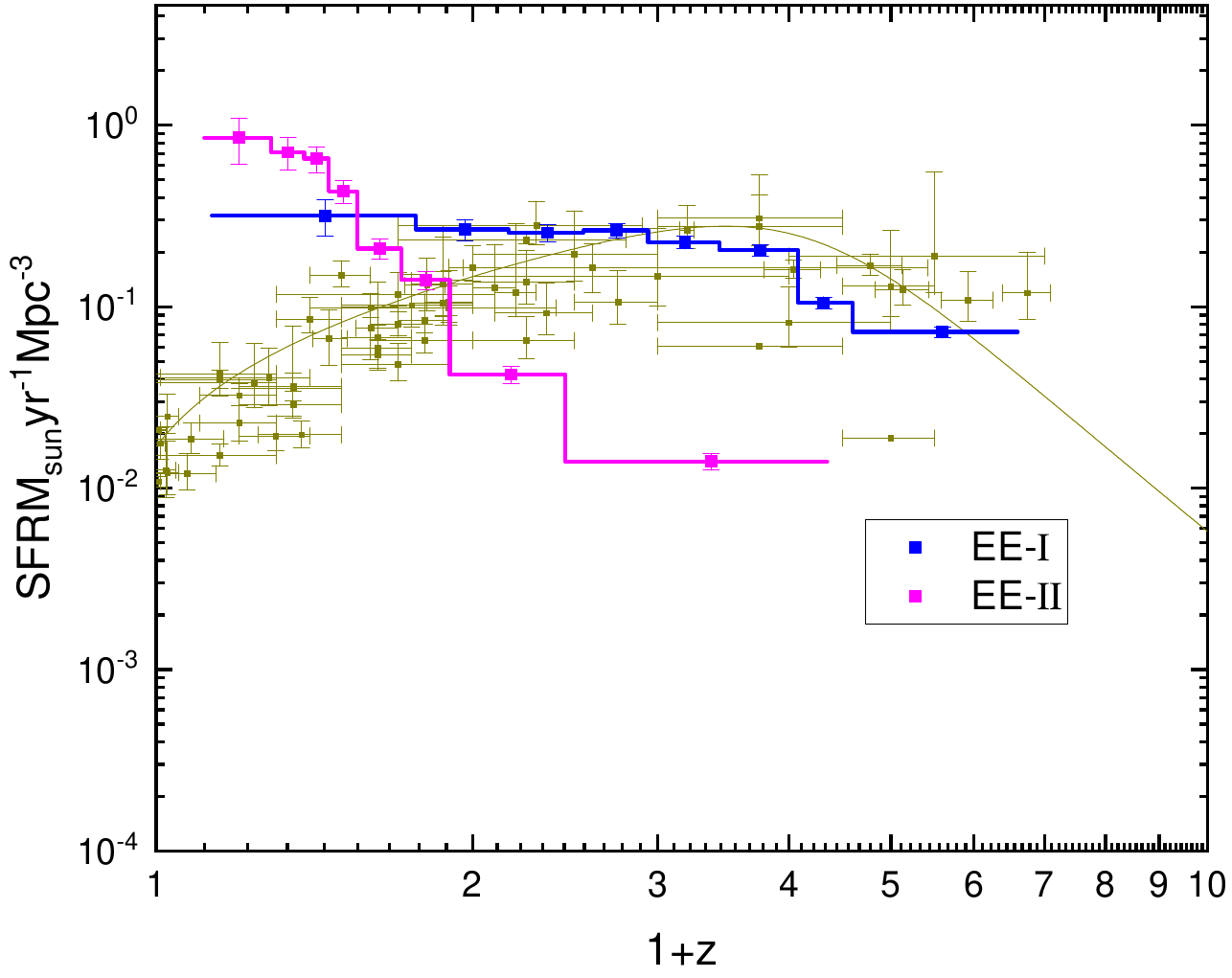}
		\caption{The formation rate for 56 EE-I and 24 EE-II GRBs. The datas of SFR is taken from \citet{2006ApJ...651..142H}.}
		\label{fig9}
	\end{figure}
	
	Fig. \ref{fig9} shows the SFR compared with the FR of EE-I and EE-II GRBs. From the qualitative perspective, The FR of EE-I GRB is roughly consistent with SFR at $z>1$. There is still fierce debate regarding the excess of $z<1$. Possible reasons include completeness, unclear definition of $T_{90}$, different origins of high- and low- luminosity bursts, etc \citep{2016A&A...587A..40P,2022MNRAS.513.1078D,2023arXiv230511380D}.

	Table \ref{tab:3} lists samples associated with supernovae and kilonovae (reference \citealp{2023MNRAS.tmp.1633L}). The GRB associated with supernovae or kilonovae is believed to originate from the death of massive stars and the merger of binary compact objects, respectively. According to the statistical results, there are a total of 6 supernova GRBs and 5 kilonovae GRB with EE. It is worth emphasizing that EE not only exists in Type I GRB (Compact Star) but also in Type II GRB (Massive Star) \citep{2020ApJ...897..154L}, and it is not a unique process for a particular type of GRB. It is a generic process that commonly exists in these two kinds of GRBs.

	\section{Conclusion} \label{sec:Conclusion}
	Gamma-ray bursts are brief and violent gamma-ray explosions in the universe, lasting from a few milliseconds to a few thousand seconds. GRBs are essential tools for tracing star formation history and studying the merger of compact objects. Using non-parametric methods to investigate the LF and FR can better understand the intrinsic properties of GRB with EE, because the co-evolution between redshift and luminosity can be removed.
	
	In this work, we used for the first time 80 GRBs with EE (56 EE-I and 24 EE-II) that they have known redshifts and well-measured spectra parameters to drive the bolometric luminosity. Then, we used a non-parametric method to derive the isotropic luminosity and formation rate based on the constructed $L-z$ plane. Before that, we had addressed the flux-truncation effect by the Kendall $\tau$ method. In our analysis, the evolution function $g(z) = {(1 + z)^k}$ can transform $L$ into $L_0$. The normalized luminosity distribution can be fitted by a broken power law after removing the redshift dependence. This fitted form can be expressed as $\psi ({L_0}) \propto L_0^{ - 0.34{\rm{ \pm 0}}{\rm{.01}}}$ for the dim segment and $\psi ({L_0}) \propto L_0^{ - 0.67{\rm{ \pm 0}}{\rm{.02}}}$ for bright segment with EE-I GRB, the broken point is ${L_0^b = 1.00 \times {{10}^{51}}erg/s}$. The form can be expressed as $\psi ({L_0}) \propto L_0^{ - 0.43{\rm{ \pm 0}}{\rm{.02}}}$ for dim segment and $\psi ({L_0}) \propto L_0^{ - 0.93{\rm{ \pm 0}}{\rm{.01}}}$  for bright segment with EE-II GRB, the broken point is ${L_0^b = 3.03 \times {{10}^{49}}erg/s}$. 
	
	We also found that the FR of the GRB EE subclass keeps decreasing, and that a broken power law can fit it. The fitting function of EE-I GRBs can be written as $\rho (z) \propto {(1 + z)^{ - 0.34{\rm{ \pm 0}}{\rm{.04}}}}$ for $z<2.39$ and $\rho (z) \propto {(1 + z)^{ - 2.34{\rm{ \pm 0}}{\rm{.24}}}}$ for $z>2.39$. 
	The FR of EE-II can describe as $\rho (z) \propto {(1 + z)^{ -1.05{\rm{ \pm 0}}{\rm{.03}}}}$ for $z<0.43$ and $\rho (z) \propto {(1 + z)^{ -8.44{\rm{ \pm 1}}{\rm{.10}}}}$ for $z>0.43$. Using Eq.\ref{equ:11}, The local formation rate are $\rho ({\rm{0) = 0.03\, Gp}}{{\rm{c}}^{{\rm{ - 3}}}}{\rm{y}}{{\rm{r}}^{{\rm{ - 1}}}}$ for EE-I GRBs and $\rho ({\rm{0) = 0.32 \,Gp}}{{\rm{c}}^{{\rm{ - 3}}}}{\rm{y}}{{\rm{r}}^{{\rm{ - 1}}}}$ for EE-II GRBs. It can be found that the formation rates of EE-I and EE-II GRBs are significantly different (see Fig. \ref{fig8}), which further suggests that these two kinds of GRB may have different origins. It is worth noting that it is difficult to search for EE-II GRBs at high redshifts because they are weaker in luminosity than EE-I GRBs, and current instruments are not sensitive to EE-II GRBs above $z > 1$. This leads to a relatively small number of EE-II GRBs at high redshift. Therefore, our result have speculative to derive a reliable FR for EE-II GRBs at $z>1$. When the sensitivity of instruments is further improved, The results would be more reliable if a sample of high redshift E-II bursts is adopted in the future.
	
	Since EE-I GRBs have a similar position in the Amati relation to lGRBs, which are thought to originate from core collapse, we further compare the formation rate of EE-I GRBs with the SFR. The results of the comparison show that the evolution of the formation rate of EE-I GRBs is similar to that of the SFR (see Fig. \ref{fig9}), suggesting that EE-I GRBs may arise from the death of massive stars, whereas EE-II GRBs, which are unrelated to the SFR, may come from other processes unrelated to the SFR. Therefore, we suggest the GRB with EE could have multiple production channels from the perspective of their formation rates.   
	
	\section{Acknowledgements}
	This work was supported by the basic research project of Yunnan Province (Grant No. 202301AT070352). The data in this article are acquired in the article \citet{2015ApJS..218...13Y} and \citet{2018ApJ...852....1Z}, Gamma-ray Coordinates Network (GCN) at \url{https://gcn.gsfc.nasa.gov/}, \url{https://www.mpe.mpg.de/ jcg/grbgen.html} and The official Swift website \url{ https://swift.gsfc.nasa.gov/results/batgrbcat/}.

	\section*{Data Availability}
	The corresponding author will share all other data underlying this article on reasonable request.

	
	
	
	\bibliographystyle{mnras}
	\bibliography{citation_list} 
	
	\onecolumn
	\begin{center}
		\renewcommand\thetable{1}
		\label{tab:1}
		{
			\tiny
			\begin{longtable}{llllllllllll}

				\caption{spectral Parameters of GRBs with EE}\label{tab:3}\\
				
				\toprule
				GRB   &$T_{90}$  & z      & $\alpha$  & $\beta$   & $E_p$            & P                  &$ S_{\gamma} $        & ${E_{min}} - {E_{max}}$ & $L_p$    &$E_{iso}$  & Type \\		
				& $s$        &        &           &           & $keV$              & $photon/cm^2/s$                   &  $ \times 10^{-7}erg/cm^2/s$                     &$keV $                        &  $\times 10^{51}erg/s$     &  $\times 10^{52}erg$       &   \\
				(1) & (2)&(3)&(4)&(5)&(6)&(7)&(8)&(9)&(10)&(11)&(12)  \\
				\midrule
				\endfirsthead
				\multicolumn{12}{c}
				{{\bfseries{Table \thetable\ }continued from previous page}} \\
				\toprule
				GRB   &$T_{90}$  & z      & $\alpha$  & $\beta$   & $E_p$            & P                  &$ S_{\gamma} $        & ${E_{min }} - {E_{max}}$ & $L_p$    &$E_{iso}$  & Type \\	
				& s        &        &           &                         & keV             &  $ photon/cm^2/s$                 &   $\times 10^{-7} erg/cm^2/s$                    &keV                         &  $ \times 10^{51}erg/s$       &  $\times 10^{52}erg$        &   \\		
				(1) & (2)&(3)&(4)&(5)&(6)&(7)&(8)&(9)&(10)&(11)&(12) \\
				\midrule
				\endhead
				\midrule
				\multicolumn{12}{c}
				{{ }} \\
				\endfoot
				\endlastfoot
				050803&87.9&0.422&-1.356&-&952.765&0.96$\pm$0.11&21.50$\pm$1.35&15-150&0.26$\pm$0.03&0.51$\pm$0.03& EE-II 
				\\
				050813&0.6&0.722&-1.19&-&64.615&1.22$\pm$0.26&1.24$\pm$0.46&15-350&0.28$\pm$0.06&0.02$\pm$0.01& EE-II\\
				050915A&52&2.5273&-1.002&-&137.16&0.77$\pm$0.14&8.50$\pm$0.88&15-150&5.56$\pm$1.01&2.19$\pm$0.23&EE-I \\
				051109A&37.2&2.346&-1.25&-&161&3.94$\pm$0.69&22.00$\pm$2.72&20-500&27.76$\pm$4.86&3.65$\pm$0.45&EE-I \\
				051111&47&1.55&-1.137&-&258.32&2.66$\pm$0.21&40.80$\pm$1.34&15-150&8.68$\pm$0.69&6.37$\pm$0.21&EE-I \\
				051221A&1.4&0.547&-1.08&-&402&4.70$\pm$0.80&24.00$\pm$4.00&20-2000&1.42$\pm$0.24&0.20$\pm$0.03& EE-II\\
				060418&103.1&1.49&-1.553&-&182.814&6.52$\pm$0.35&83.30$\pm$2.53&15-150&15.53$\pm$0.83&11.34$\pm$0.34&EE-I \\
				060502B&0.131&0.287&-0.11&-&117.949&0.62$\pm$0.12&0.40$\pm$0.05&15-150&0.02$\pm$0.00&0.00$\pm$0.00& EE-II\\
				060526&298.2&3.21&-0.336&-&89.792&1.67$\pm$0.18&12.60$\pm$1.65&15-150&16.02$\pm$1.73&3.55$\pm$0.46&EE-I \\
				060607A&102.2&3.082&-1.147&-&149.843&1.40$\pm$0.13&25.50$\pm$1.12&15-150&17.22$\pm$1.60&9.92$\pm$0.44&EE-I \\
				060614$^*$&108.7&0.13&-2.037&-&98.52&11.50$\pm$0.74&204.00$\pm$3.63&15-150&0.24$\pm$0.02&0.61$\pm$0.01&EE-I \\
				060708&10.2&1.92$^b$&-1.22&-&88.948&1.94$\pm$0.14&4.94$\pm$0.37&15-150&5.90$\pm$0.43&0.73$\pm$0.05&EE-I \\
				060719&66.9&1.532&-1.63&-&68.556&2.16$\pm$0.20&15.00$\pm$0.91&15-150&4.35$\pm$0.40&1.86$\pm$0.11&EE-I \\
				060814&145.3&1.9229$^b$&-1.412&-&302.336&7.27$\pm$0.29&146.00$\pm$2.39&15-150&40.41$\pm$0.61&36.85$\pm$0.60&EE-I \\
				060904B$^c$&171.5&0.703&-1.234&-&84.094&2.44$\pm$0.21&16.20$\pm$1.43&15-150&0.61$\pm$0.05&0.34$\pm$0.03&EE-I \\
				060927&22.5&5.6&-0.7&-&70.673&2.70$\pm$0.17&11.30$\pm$0.68&15-150&85.99$\pm$5.41&7.68$\pm$0.46&EE-I \\
				061007&75.3&1.261&-0.7&-2.61&498&14.60$\pm$0.37&444.00$\pm$5.62&20-10000&37.11$\pm$0.94&18.66$\pm$0.24&EE-I \\
				061110A&40.7&0.758&-1.556&-&240.328&0.53$\pm$0.12&10.60$\pm$0.76&15-150&0.26$\pm$0.06&0.42$\pm$0.03& EE-II\\
				061201$^*$&0.21&0.111&-0.36&-&873&3.50$\pm$0.35&53.30$\pm$7.00&20-3000&0.07$\pm$0.01&0.02$\pm$0.00& EE-II\\
				070103&18.6&2.6208&-1.223&-&46.633&1.04$\pm$0.15&3.38$\pm$0.46&15-150&5.62$\pm$0.81&0.83$\pm$0.11&EE-I \\
				070208&47.7&1.165&-1.651&-&51.29&0.90$\pm$0.22&4.45$\pm$1.01&15-150&0.88$\pm$0.22&0.33$\pm$0.07&EE-I \\
				070429B&0.47&0.904&-1.099&-&72.852&1.76$\pm$0.24&0.63$\pm$0.10&15-150&0.74$\pm$0.10&0.02$\pm$0.00& EE-II\\
				070612A&368.8&0.617&-1.439&-&137.695&1.51$\pm$0.38&106.00$\pm$6.01&15-150&0.35$\pm$0.09&2.16$\pm$0.12&EE-I \\
				070721B&340&3.626&-0.41&-&224.25&1.50$\pm$0.30&36.00$\pm$2.00&15-150&45.63$\pm$9.13&23.62$\pm$1.31&EE-I \\
				070724A&0.4&0.457&-1.15&-&82&0.94$\pm$0.09&0.30$\pm$0.03&15-150&0.08$\pm$0.01&0.00$\pm$0.00& EE-II\\
				071227&1.8&0.383&-0.7&-&1000&1.68$\pm$0.17&16.00$\pm$2.00&20-13000&0.50$\pm$0.05&0.06$\pm$0.01& EE-II\\
				080310&365&2.4266&-1.65&-&23.284&1.30$\pm$0.20&23.00$\pm$2.00&15-150&7.71$\pm$1.19&7.39$\pm$0.64&EE-I \\
				080413A&46&2.433&-1.2&-&170&5.60$\pm$0.20&35.00$\pm$1.00&15-1000&36.51$\pm$1.30&5.44$\pm$0.16&EE-I \\
				080413B&8&1.1&-1.26&-&73.3&18.70$\pm$0.80&32.00$\pm$1.00&15-150&13.54$\pm$0.58&1.63$\pm$0.05&EE-I \\
				080430&16.2&0.767&-1.645&-&151.587&2.60$\pm$0.20&12.00$\pm$1.00&15-150&1.15$\pm$0.09&0.44$\pm$0.04&EE-I \\
				080603B&60&2.69&-1.21&-&71&3.50$\pm$0.20&24.00$\pm$1.00&15-150&22.24$\pm$1.27&6.13$\pm$0.26&EE-I \\
				080607&79&3.036&-0.76&-2.57&348&23.10$\pm$1.10&240.00$\pm$0.00&20-4000&380.20$\pm$18.10&49.45$\pm$0.00&EE-I \\
				080707&27.1&1.23&-1.397&-&29.067&1.00$\pm$0.10&5.20$\pm$0.60&15-150&0.90$\pm$0.09&0.39$\pm$0.04 &EE-I \\
				080810&106&3.35&-2.5&-1.2&2523&2.00$\pm$0.20&46.00$\pm$2.00&15-1000&115.94$\pm$6.13&19.50$\pm$0.08& EE-II\\
				080905A&1&0.1218&0.12&-2.35&311.2&6.32$\pm$0.20&8.51$\pm$0.20&10-1000&0.05$\pm$0.00&0.00$\pm$0.00& EE-II\\
				080905B&128&2.374&-1.579&-&256.097&0.50$\pm$0.10&18.00$\pm$2.00&15-150&4.27$\pm$0.85&6.43$\pm$0.71&EE-I \\
				090205&8.8&4.7&-0.394&-&38.42&0.50$\pm$0.10&1.90$\pm$0.30&15-150&8.11$\pm$1.62&19.50$\pm$0.15&EE-I \\
				090407&310&1.4485&-1.585&-&309.959&0.60$\pm$0.10&11.00$\pm$2.00&15-150&1.63$\pm$0.27&1.71$\pm$0.31& EE-II\\
				090424&48&0.544&-1.19&-&108.6&71.00$\pm$2.00&210.00$\pm$0.00&15-150&10.47$\pm$0.29&2.74$\pm$0.00&EE-I \\
				090426A&1.2&2.609&-1.105&-&55.08&2.40$\pm$0.30&1.80$\pm$0.30&15-150&12.61$\pm$1.58&0.41$\pm$0.07&EE-I \\
				090510&0.3&0.903&-0.86&-2.58&4302&40.95$\pm$4.10&33.70$\pm$3.40&10-1000&210.78$\pm$21.10&0.72$\pm$0.07& EE-II\\
				090530&48&1.266&-1.078&-&92.142&2.50$\pm$0.30&11.00$\pm$1.00&15-150&2.66$\pm$0.32&0.71$\pm$0.06&EE-I \\
				090618$^c$&113.2&0.54&-1.42&-&134&38.90$\pm$0.80&1050.00$\pm$10.00&15-150&6.49$\pm$0.13&15.94$\pm$0.15&EE-I \\
				090715B&266&3&-1.1&-&134&3.80$\pm$0.20&57.00$\pm$2.00&20-2000&43.63$\pm$2.30&13.10$\pm$0.46&EE-I \\
				100425A&37&1.755&-0.885&-&25.353&1.40$\pm$0.20&4.70$\pm$0.90&15-150&2.44$\pm$0.35&0.62$\pm$0.12&EE-I \\
				100704A&197.5&3.6&-0.76&-2.53&178.3&4.30$\pm$0.20&60.00$\pm$2.00&10-1000&53.66$\pm$2.50&16.30$\pm$0.54&EE-I \\
				100724&1.4&1.288&-0.51&-&42.5&1.56$\pm$0.16&1.41$\pm$0.14&15-150&1.15$\pm$0.12&0.08$\pm$0.01&EE-I \\
				100728A&198.5&1.567&-0.76&-&357.7&5.10$\pm$0.20&380.00$\pm$0.00&50-300&37.53$\pm$1.47&51.73$\pm$0.00&EE-I \\
				100906A&114.4&1.727&-1.34&-1.98&106&10.10$\pm$0.40&120.00$\pm$0.00&50-300&78.21$\pm$205.15&28.71$\pm$0.91&EE-I \\
				110715A&13&0.82&-1.23&-2.7&120&53.90$\pm$1.10&118.00$\pm$2.00&20-10000&22.23$\pm$125.09&2.73$\pm$10.40&EE-I \\
				111008A&63.46&5&-1.36&-&149&6.40$\pm$0.70&53.00$\pm$3.00&20-2000&275.27$\pm$30.11&29.35$\pm$1.66&EE-I \\
				111228A$^c$&101.2&0.71627&-1.9&-&34&12.40$\pm$0.50&85.00$\pm$2.00&50-300&17.92$\pm$0.72&4.43$\pm$0.10&EE-I \\
				120729A$^c$&71.5&0.8&-0.78&-&64.985&2.90$\pm$0.20&24.00$\pm$1.00&15-150&0.80$\pm$0.06&0.54$\pm$0.02&EE-I \\
				121027A&62.6&1.773&-1.58&-&82.46&1.30$\pm$0.20&20.00$\pm$1.00&15-150&3.77$\pm$0.58&3.19$\pm$0.16&EE-I \\
				130427A$^c$&162.83&0.3399&-0.789&-3.06&830&331.00$\pm$4.60&3100.00$\pm$30.00&10-1000&46.60$\pm$0.65&8.98$\pm$0.09&EE-I \\
				130514A&204&3.6&-1.44&-2.5&108&2.80$\pm$0.30&91.00$\pm$2.00&15-1200&36.62$\pm$3.92&33.52$\pm$0.74&EE-I \\
				130603B$^*$&0.18&0.356&-0.73&-&660&1.30$\pm$0.20&66.00$\pm$7.00&20-15000&0.23$\pm$0.04&0.21$\pm$0.02& EE-II\\
				130907A&>360&1.238&-0.91&-2.42&394&25.60$\pm$0.50&1400.00$\pm$10.00&20-10000&47.45$\pm$0.93&58.55$\pm$0.42&EE-I \\
				131004&1.54&0.71&-1.36&-22.09&118.1&9.82$\pm$0.98&5.09$\pm$0.51&10-1000&2.09$\pm$0.21&0.08$\pm$0.01& EE-II\\
				140430A&173.6&1.6&-2.108&-&47.62&2.50$\pm$0.20&11.00$\pm$2.00&15-150&514.51$\pm$41.16&141.31$\pm$25.69&EE-I \\
				140506A&111.1&0.889&-0.9&-2&141&10.90$\pm$0.90&28.00$\pm$3.00&50-300&12.41$\pm$1.02&0.95$\pm$0.10&EE-I\\
				140903A$^*$&0.3&0.351&-1.36&-&44.169&2.50$\pm$0.20&1.40$\pm$0.10&15-150&0.10$\pm$0.01&0.01$\pm$0.00& EE-II\\
				141004A$^c$&3.92&0.57&-1.3&-&147&6.10$\pm$0.30&6.70$\pm$0.30&50-300&2.85$\pm$0.14&0.11$\pm$0.01& EE-II\\
				141212A&0.3&0.569&-1.146&-&94.865&1.20$\pm$0.20&0.72$\pm$0.12&15-150&0.18$\pm$0.03&0.01$\pm$0.00& EE-II\\
				150120A&1.2&0.46&-1.43&-1.65&130&3.10$\pm$0.30&3.40$\pm$0.80&10-1000&0.42$\pm$0.02&0.04$\pm$0.01& EE-II\\
				150423A&0.22&1.394&0.43&-&120&0.90$\pm$0.10&0.63$\pm$0.10&15-150&1.55$\pm$0.17&0.04$\pm$0.01& EE-II\\
				151027A&129.69&0.81&-1.41&-&340&6.80$\pm$0.60&78.00$\pm$2.00&50-300&10.36$\pm$0.91&3.47$\pm$0.09&EE-I \\
				160227A&316.5&2.38&-0.75&-&65.8&0.60$\pm$0.10&31.00$\pm$2.00&15-150&2.42$\pm$0.40&5.38$\pm$0.35&EE-I \\
				160410A&8.2&1.717&-0.71&-&1416&3.50$\pm$0.30&7.80$\pm$0.80&20-10000&51.14$\pm$4.38&0.54$\pm$0.06& EE-II\\
				160425A&304.58&0.555&-1.975&-&5.251&2.80$\pm$0.20&21.00$\pm$2.00&15-150&0.56$\pm$0.04&0.48$\pm$0.05&EE-I \\
				160624A&0.2&0.483&-0.63&-3.65&1168&6.39$\pm$0.64&1.21$\pm$0.12&10-1000&3.35$\pm$0.41&0.01$\pm$0.00& EE-II\\
				160824B&0.48&0.16&-0.12&5.38&46.32&1.68$\pm$0.17&1.03$\pm$0.10&15-150&0.01$\pm$0.00&0.00$\pm$0.00& EE-II\\
				161017A&216.3&2.0127&-1.04&-&298.5&2.80$\pm$0.20&53.00$\pm$2.00&50-300&34.32$\pm$2.45&10.89$\pm$0.41&EE-I \\
				161108A&105.1&1.159&-1.312&-&65.124&0.60$\pm$0.10&11.00$\pm$1.00&15-150&0.49$\pm$0.08&0.63$\pm$0.06&EE-I \\
				170705A&217.3&2.01&-0.88&-2.38&100&13.90$\pm$0.40&95.00$\pm$3.00&50-300&104.92$\pm$3.02&18.62$\pm$0.59&EE-I \\
				180329B&210&1.998&-0.97&-&48.6&1.40$\pm$0.40&33.00$\pm$3.00&15-150&3.54$\pm$1.01&4.53$\pm$0.41&EE-I \\
				180620B&198.8&1.1175&-0.85&-&149&3.60$\pm$0.20&100.00$\pm$3.00&15-150&3.74$\pm$0.21&5.88$\pm$0.18&EE-I \\
				190719C&185.7&2.469&-0.87&-&81&5.50$\pm$0.30&51.00$\pm$3.00&50-300&75.14$\pm$4.10&13.17$\pm$0.77&EE-I \\
				200522A$^*$&0.62&0.554&-0.54&-&77.76&1.50$\pm$0.20&1.10$\pm$0.10&15-150&0.08$\pm$0.01&0.01$\pm$0.00& EE-II\\
				210619B&60.9&1.937&-1.25&-&286.165&115.00$\pm$2.20&950.00$\pm$10.00&15-150&666.57$\pm$12.75&235.66$\pm$2.48&EE-I \\
				
				\bottomrule
			\end{longtable}
		}
		\footnotesize{Note: a:The spectral parameters in this article are acquired in the article \citet{2015ApJS..218...13Y} and \citet{2018ApJ...852....1Z}, Gamma-ray Coordinates Network (GCN) at \url{https://gcn.gsfc.nasa.gov/}, \url{https://www.mpe.mpg.de/ jcg/grbgen.html} and The official Swift website \url{ https://swift.gsfc.nasa.gov/results/batgrbcat/}\\
			b: The values of redshift of GRB 060708 and GRB 060814 come from \citet{2012grb..confE.136H}\\
			c: The GRB is associated with supernovae.\\
			*: The GRB is associated with kilonovae.}

	\end{center}



	\bsp	
	\label{lastpage}
\end{document}